\documentclass[aps,twocolumn,nofootinbib]{revtex4-2}

\usepackage{graphicx}
\usepackage{physics}
\usepackage{hyperref}
\usepackage{cleveref}

\graphicspath{{Fig/}}

\usepackage{amsthm}

\usepackage{amsmath}

\begin{document}

\title{Dopant site occupancy determined by core-loss-filtered, position-averaged convergent beam electron diffraction}

\author{Michael Deimetry}
\author{Timothy C. Petersen}
\author{Matthew Weyland}
\author{Scott D. Findlay\thanks{Corresponding author: \href{mailto:scott.findlay@monash.edu}{scott.findlay@monash.edu}}}

\affiliation{School of Physics and Astronomy, Monash University, Wellington Rd, Victoria 3800, Australia}
\affiliation{Monash Centre for Electron Microscopy, Monash University, Wellington Rd, Victoria 3800, Australia}
\affiliation{Department of Materials Science \& Engineering, Monash University, Wellington Rd, Victoria 3800, Australia}

\date{\today}

\begin{abstract}
In the elastic scattering regime, probe position-averaged convergent beam electron diffraction (PACBED) patterns have proven robust for estimating specimen thickness and mistilt. Through simulation, we show that core-loss-filtered PACBED patterns can be used to measure the site occupancy of a small concentration of dopants in an otherwise known crystal structure. By leveraging the reciprocity between scanning and conventional transmission electron microscopy, we interpret core-loss PACBED patterns using a strategy traditionally used for determining dopant concentrations via energy dispersive X-ray spectroscopy. We show that differences in the interaction range of different elements hinder a purely measurement-based quantification strategy, but that this can be overcome through comparison with simulations that generalize the Cliff-Lorimer $k$-factors.
\end{abstract}

\maketitle

\section{Introduction}\label{sec:introduction}
Fast read-out direct electron detectors have facilitated the acquisition of 4D scanning transmission electron microscopy (4D STEM) datasets wherein the momentum distribution of scattered electrons in the diffraction pattern is recorded as a function of a 2D raster scan of a convergent electron probe \citep{10.1017/S1431927619000497}. This capability can be further refined through energy filtering, which selectively detects electrons based on their energy loss. Zero-loss filtering 4D STEM data can improve the signal-to-background and reliability of analyses that are based on elastic scattering alone \citep{bustillo20214d,chejarla2023measuring,li2024efficient,gebhart2025grain}. Filtering the low loss scattering can be used to probe dispersion relations \citep{qi2021four} and explore the angular scattering distribution of plasmons \citep{robert2022contribution}. That core-loss-filtered convergent beam electron diffraction patterns may contain rich structure and so potentially useful information has long been anticipated \citep{midgley1995energy,muller2017measurement}, and methods for simulating it are known \citep{PhysRevB.77.184107}. However, there have been very few published attempts at acquiring core-loss-filtered convergent beam electron diffraction patterns \citep{hass_koch}, and only a few proposals for how such data might usefully be interpreted \citep{loffler20234d,DEIMETRY2024114036}.

This paper considers the potential application of core-loss-filtered position averaged convergent beam electron diffraction (PACBED), which collapses the 4D dataset by averaging over all probe positions. Zero-loss PACBED, as proposed by \citet{PACBED_lebeau}, has proven fruitful for determining crystal thickness and tilt, but in the core-loss regime sensitivity to thickness is lost due to the large inelastic mean free path. However, via (approximate) reciprocity, we show that core-loss-filtered PACBED patterns are essentially the same as two-dimensional rocking patterns for electron energy loss spectroscopy (EELS), whereby the signal from a select EELS edge is plotted as a function of the orientation of an incident plane-wave electron beam. It is well known that rocking patterns, albeit more usually those from energy-dispersive X-ray spectroscopy (EDX) than from EELS, are characteristic of the atomic crystallographic distribution of the species of the corresponding core loss: this is the basis of the atom location by channelling enhanced microanalysis (ALCHEMI) technique\footnote{There are multiple terms for rocking-curve-like experiments, including electron channelling patterns (ECPs, more often used in scanning electron microscopy and/or using back-scattered electrons) \citep{joy1982electron}, high angular resolution electron channelling X-ray spectroscopy (HARECXS) \citep{matsumura1999electron}, and high angular resolution electron channelling electron spectroscopy (HARECES) \citep{zaluzec2005hareces}. We adopt the term ALCHEMI here since we share that technique's goal of locating atoms through the channelling (dynamical diffraction) enhanced detail in patterns formed using spectroscopic signals. Following \citet{buseck1992electron}, we use the terms EDX ALCHEMI and EELS ALCHEMI when it is needful to clarify the signal used.} for seeking to quantify the site occupancy of a relatively low concentration of dopant atoms \citep{spence_alchemi_1983,buseck1992electron,rossouw_statistical_alchemi,rossouw1996statistical,jones2003determining,muto2017high,ohtsuka20212d}.

The traditional ALCHEMI literature established that localization of the interaction is important for quantifying the site occupancy of the dopants, especially via analysis from the measured data alone, and hence that EDX signals are more reliable than the EELS signals \citep{spence1988localization,buseck1992electron,jones2003determining,ofori2005determining}. Nevertheless, recognising that core-loss-filtered PACBED provides a new route to acquiring EELS ALCHEMI-like data, in this paper we revisit the limitations of less localized interactions in EELS ALCHEMI. Building on the work of \citet{OXLEY1999109}, we show how these limitations can be overcome through comparison with simulations for the case of modest dopant concentrations in an otherwise known host structure. We further discuss the spatial field of view on which core-loss-filtered PACBED can reliably be performed, which is constrained by the variability in signal due to the particular configuration of dopants and by the dose needed for adequate signal above noise.

\section{Core-loss-filtered PACBED and reciprocity}\label{sec:PACBED_reciprocity}

\begin{figure*}
    \centering
    \includegraphics[width=0.95\linewidth]{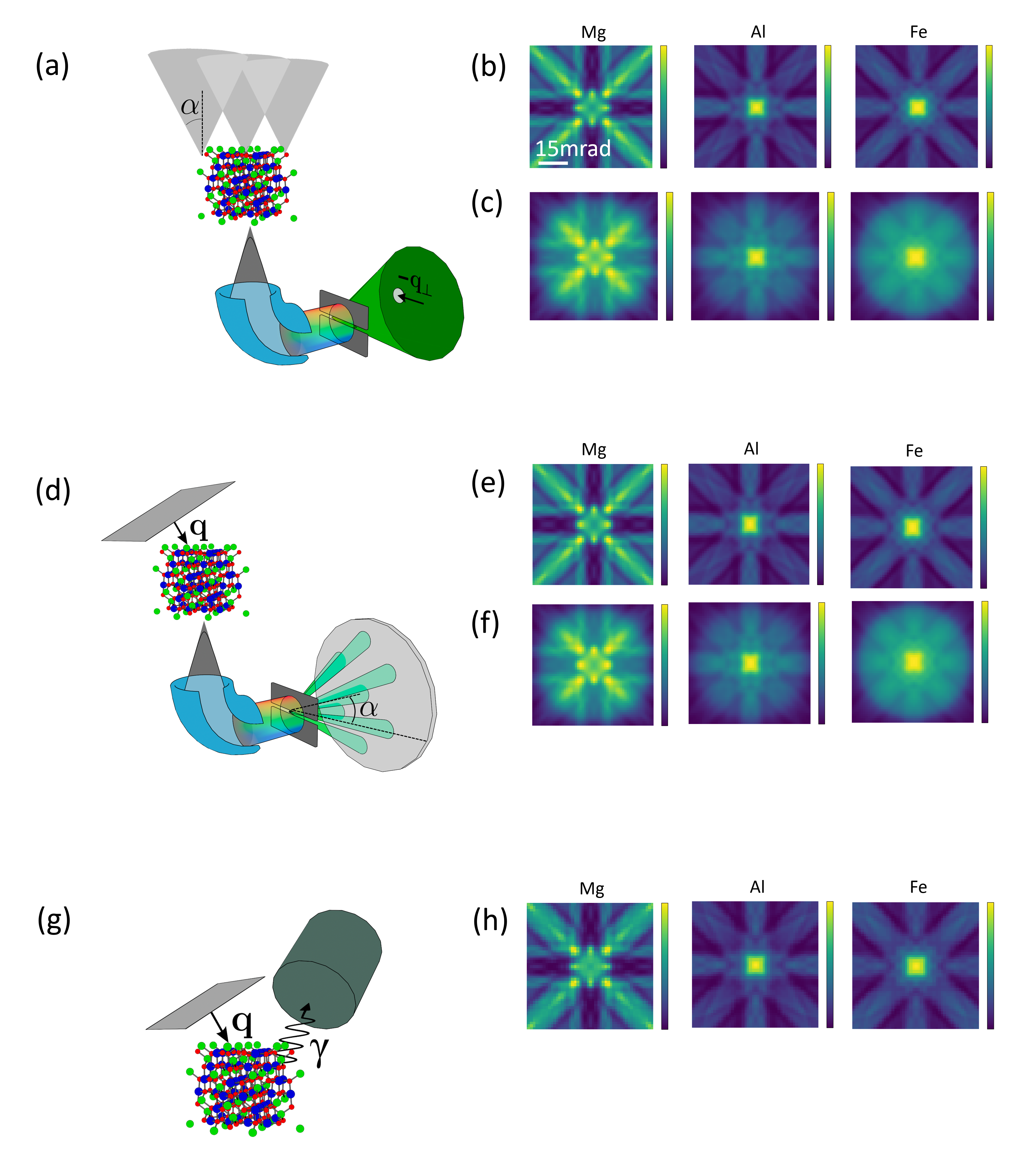}
    \caption{(a) Core-loss-filtered PACBED experiment wherein a 4D-STEM dataset is averaged over scan positions, with the detector pixel at $-{\bf q}_\perp$ indicated, together with corresponding simulated images for (b) $\alpha=65$mrad and (c) $\alpha=30$mrad probe forming apertures for an iron-doped spinel with Mg-K, Al-K, Fe-L$_1$ ionization and assuming a narrow energy window 10eV above the corresponding edge. (d) Rocking curve CTEM experiment for an incident plane-wave where signal on an EELS detector with semiangle $\alpha$ is recorded as a function of plane-wave illumination wavevector ${\bf q}=({\bf q}_\perp,q_z)$ direction, together with corresponding simulated incoherent channelling patterns (ICPs) for (e) $\alpha=65$mrad and (f) $\alpha=30$mrad, again for an iron-doped spinel with Mg-K, Al-K, Fe-L$_1$ ionization and assuming a narrow energy window 10eV above the corresponding edge. The approximate reciprocity between (a) and (d) explains why the patterns in (b) and (e) are indistinguishable, and likewise the patterns in (c) and (f). (g) Rocking curve CTEM experiment where X-ray (denoted with $\gamma$) signal on an EDX detector is recorded as a function of plane-wave illumination incidence angle. (h) Corresponding simulated ICPs, with very similar appearance to the patterns in (b) and (e).}
    \label{fig:ICPs_showcase}
\end{figure*}

\Cref{fig:ICPs_showcase}(a) shows the simplified experimental setup for core-loss-filtered PACBED. With the detector behind the energy filter, a diffraction pattern can be formed from the electrons that have energy loss within some chosen window beyond the characteristic ionization energy of one of the core shells of one of the atomic species present. The energy-filtered diffraction pattern is then averaged over probe positions, either by continuously acquiring during the scan or else by acquiring a core-loss-filtered 4D STEM data set and averaging over probe positions in post-processing. Throughout this manuscript, we will use the specific case study of the spinel structure MgAl$_2$O$_4$, doped with Fe on Mg and/or Al sites. This material was chosen due to the materials science interest in its properties when doped \citep{LEE2001376,aizawa2002characteristics,MARIOSI20202772}, and because its structure allows dopants to be distributed across distinct crystallographic sites. Assuming 5\% and 7\% Fe occupancy on Mg and Al sites respectively (note: all such percentages in this paper are atomic percentages) and a 1eV energy window situated 10eV above the ionization threshold for each edge, \cref{fig:ICPs_showcase}(b) and (c) show simulated core-loss PACBED patterns for the Mg-K shell, Al-K shell and Fe-L$_1$ shell. The patterns in \cref{fig:ICPs_showcase}(b) assume a 65mrad probe-forming aperture semiangle while those in \cref{fig:ICPs_showcase}(c) assume a 30mrad probe-forming aperture semiangle. In \cref{fig:ICPs_showcase}(c), the intensity fall-off outside the bright field disk is evident; in \cref{fig:ICPs_showcase}(b) only the central portion of the bright field region at that aperture size is shown. The Mg and Al patterns clearly have different structure, reflecting the different crystallographic sites of those species. The Fe pattern looks more like the Al pattern than the Mg pattern, reflecting the higher number of Fe atoms on Al sites than on Mg sites in this example.

We gain some insight into how the information in these patterns could be used by recognising that this imaging mode is, via reciprocity, equivalent to a much-studied imaging mode in conventional TEM (CTEM). As noted by \citet{cowley_1969}, some commonality between CTEM and STEM imaging is due to the reciprocal geometries in conjunction with a scattering reciprocity which dictates that the exchange of a point source and detector yield identical measurements. Strictly, reciprocity from core-loss scattering would require the crystal to begin in an excited state which then de-excites by imparting energy to the beam electron, a circumstance not feasible to arrange. However, provided the energy loss is small compared to the accelerating voltage, an approximate reciprocity holds for a crystal initially in the ground state \citep{pogany_turner,KOHL1985173,FINDLAY200758,KRAUSE20171}. \citet{FINDLAY200758} have shown that
 \begin{align}
     \Psi^{\text{STEM}}({\bf r}_0, -{\bf q}_\perp) \approx \Psi^{\text{CTEM}}({\bf r}_0, {\bf q}_\perp) 
     \label{eq:receprocity_eqn}
 \end{align}
 where $\Psi$ denotes an inelastic wavefunction, ${\bf r}_0$ denotes the probe position in STEM but the (real-space) image coordinate in CTEM, and ${\bf q}_\perp$ denotes a reciprocal space location in the detector plane in STEM but the transverse component of wave vector of an incident plane-wave in CTEM. There are a number of subtleties in this approximate equality, including the sign difference between the ${\bf q}_\perp$ terms on the two sides of \cref{eq:receprocity_eqn} and  that we assume the same single transition of a single atom but the location of that atom and the orientation of the sample reflect the direction reversal of reciprocity. The interested reader is referred to \citet{FINDLAY200758} for these details.
 
 Here we only want to motivate the main idea: the approximate equality of \cref{eq:receprocity_eqn} remains valid when on both sides we take the intensity, sum over all contributing transitions (say for a particular energy loss above ionization threshold for a particular atom) and atoms, and further sum over ${\bf r}_0$ values for a repeating unit. The sum over position coordinates in CTEM gives the total intensity within the assumed image-forming aperture, which is precisely the total intensity in the CTEM diffraction pattern within a circular detector of equivalent angular extent. Thus we expect the CTEM configuration in \cref{fig:ICPs_showcase}(d), where ${\bf q}_\perp$ is the transverse component of the incident plane-wave wavevector ${\bf q}$ and $\alpha$ is the EELS detector semiangle, to be reciprocity-related to a good approximation to the STEM configuration in \cref{fig:ICPs_showcase}(a), where $-{\bf q}_\perp$ identifies the detector pixel recording the scattered intensity component with that transverse wavevector, $\alpha$ is the probe-forming aperture semiangle, and the detected intensity is understood to be averaged over probe positions. The rocking pattern simulations in \cref{fig:ICPs_showcase}(e) and (f) were calculated assuming the CTEM geometry of \cref{fig:ICPs_showcase}(d) and an EELS detector extent of 65mrad and 30mrad, respectively. That reciprocity holds to an excellent approximation is seen by \cref{fig:ICPs_showcase}(b) and (e) being indistinguishable from one another, and likewise  \cref{fig:ICPs_showcase}(c) and (f). These rocking patterns are sometimes called incoherent (or ionization) channelling patterns (ICPs) \citep{rossouw1995incoherent,rossouw1997generation,muto2017high}, and for succinctness we will adopt that term here, extending it to core-loss-filtered PACBED patterns.

The plane-wave rocking experiments traditionally used to determine doping information using ALCHEMI are not based on EELS signals but rather on EDX signals. The experimental configuration for this is depicted in \cref{fig:ICPs_showcase}(g), with corresponding ICPs shown in \cref{fig:ICPs_showcase}(h). These ICPs are very similar to those in \cref{fig:ICPs_showcase}(b) and (e) because the calculations are similar: the EDX calculation is essentially an EELS calculation that includes all scattering angles and energy losses above threshold. The similarity between \cref{fig:ICPs_showcase}(e) and (h) is thus attributed to the 65mrad detector being large enough to encompass the majority of intensity scattered by ionization events at the given energy assumed. The residual differences are attributed to the difference between considering a single energy loss above threshold and summing over all losses above threshold. Consequently, if we could acquire core-loss-filtered PACBED patterns like those in \cref{fig:ICPs_showcase}(b) they could be used instead of traditional EDX rocking ICPs for ALCHEMI analysis of dopant site occupancy. However, \cref{fig:ICPs_showcase}(b) assumes a probe-forming aperture semiangle of 65mrad. On a perfect crystal, PACBED patterns are independent of lens aberrations \citep{PACBED_lebeau} and so would not in principle be affected by an inability to balance aberrations across such a wide aperture. (Doping is not strictly periodic but, as we show later, periodicity is a good approximation provided one averages over multiple unit cells.) In practice using highly aberrated probes can be challenging, making it difficult to align the sample and identify a desired region of interest. However, we will presently show that using narrower probe-forming aperture semiangles, like the 30mrad assumed in \cref{fig:ICPs_showcase}(c), complicates the ALCHEMI analysis. Traditional ALCHEMI relies on the ICP for the dopant being a linear combination of the ICPs from the elements at the host sites. Though not immediately obvious from the figures, this proves to be a good approximation for \cref{fig:ICPs_showcase}(b), (e) and (h) but not for \cref{fig:ICPs_showcase}(c) and (f).

If this can be overcome, there are potential advantages and complementarities to the core-loss-filtered PACBED approach. First, with fast readout direct electron detectors located behind an energy filter becoming more readily available, obtaining energy-filtered PACBEDs fits readily within standard STEM acquisition strategies. By contrast, carefully controlled plane-wave rocking synchronised with EDX detection is less routine, requiring a bespoke experiment. Second, EELS count rates tend to be higher for lighter elements, whereas EDX count rates depend on the fluorescence rate (the probability of X-ray emission per ionization event) that drops with atomic number, making it harder to detect very light elements such as lithium.\footnote{
The case for detector collection efficiency is less clear cut. EDX has a low geometric collection efficiency due to the limited detector collection angle, potentially exacerbated by geometric limitations whereby the relative position and orientation of sample, holder and detector might mean signal loss due to absorption of X-rays in the sample and detector holder. Conversely, an EELS detector in the forwards direction has a much higher geometric collection efficiency. However, EDX effectively integrates over all energy losses of the fast electron, whereas the energy window width in EELS needs to be quite small to enable reliable background subtraction.} Third, acquiring 4D STEM data provides the freedom to consider different regions of the crystal in post-processing, whereas rocked plane-waves probing differently sized regions would require separate experiments with different beam footprints. Fourth, STEM offers the advantage of simultaneously acquiring multiple imaging modes such as high-angle annular dark field or EDX STEM images (if the aberrations are balanced within the probe-forming aperture chosen), which could be correlated with the energy-filtered diffraction patterns. Relative disadvantages of the core-loss-filtered PACBED approach include that the different energy losses required to effect background subtraction must be acquired sequentially (whereas plane-wave rocking experiments acquire each spectrum in parallel), and that the signal-to-background ratio in EDX is much higher than in EELS. Whether the damage mechanisms make STEM or CTEM more favourable will depend on the material of interest. 

Before considering how ALCHEMI analysis might be applied to core-loss-filtered PACBED patterns, let us first review how scattering and crystal doping are handled in simulation within the core-loss regime.

 \section{Inelastic Channels Formulation of Core-Loss Scattering}

\begin{figure*}
    \centering
    \includegraphics[width=1\linewidth]{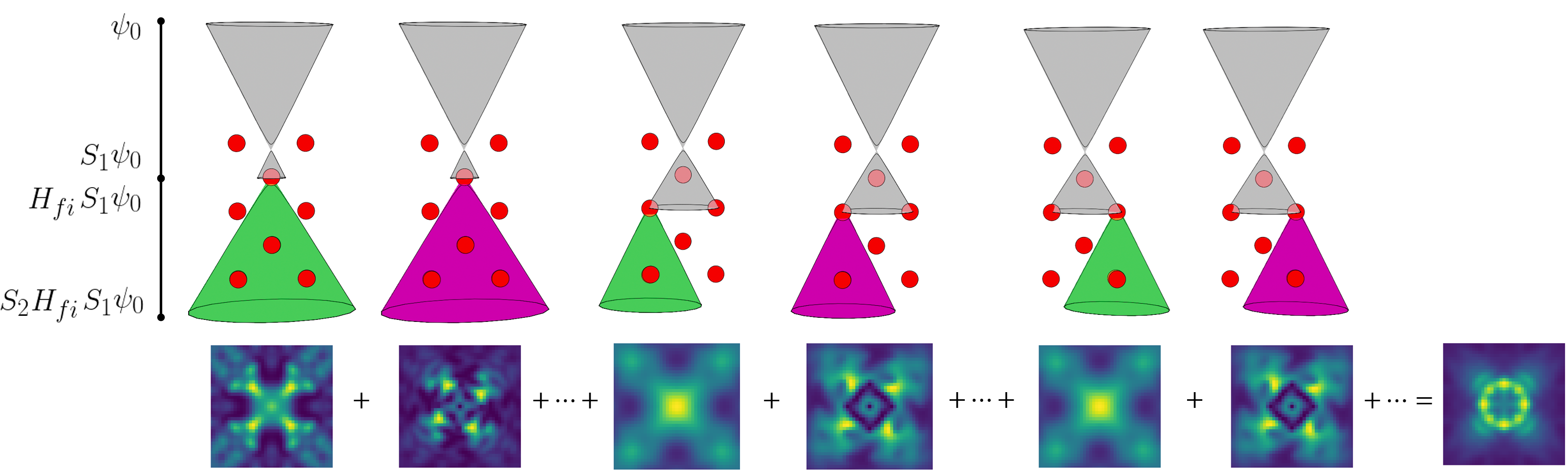}
    \caption{Process for simulating core-loss PACBED patterns as performed using $S$-matrices. The incident beam, $\psi_0$ (grey), elastically scatters through the specimen, as described by the operator $S_1$. When the wavefield encounters an atom of suitable core-shell energy, an inelastic wavefield (green or purple, indicating two of the many different transitions contributing) is generated as per \cref{eq:elastic_to_inelastic}, which is subsequently propagated elastically through the specimen using $S_2$. The various inelastic diffraction pattern intensities are added as per \cref{eq:total_intensity}. (Since the inelastic mean free path for ionization is larger than the sample thicknesses we neglect double inelastic scattering, which would anyway produce a different energy loss.)}
    \label{fig:calc_demo}
\end{figure*}

\Cref{fig:calc_demo} summarises our approach to simulating core-loss PACBED patterns. The incident wavefield (grey shading in the figure) scatters elastically through the crystal until reaching the depth of each atom being ionized. Following \citet{coene_inelastic_1990,dwyer_multislice_2005,dalfonso_three-dimensional_2008}, the inelastically scattered wavefield immediately after ionization $\psi_f(\mathbf{r}_\perp)$ is related to the wavefield immediately before ionization $\psi(\mathbf{r}_\perp)$ via 
\begin{align}
    \psi_f(\mathbf{r}_\perp) = -i\sigma H_{fi}(\mathbf{r}_\perp)\psi(\mathbf{r}_\perp) \;,
    \label{eq:elastic_to_inelastic}
\end{align}
where $\sigma$ is an interaction constant and $H_{fi}$ is a transition potential with $f$ and $i$ labelling the final and initial states in the transition considered. The inelastic wavefield then scatters elastically through the remainder of the crystal. For each ionized atom, multiple transition potentials must be considered, corresponding to different possible final states consistent with the same energy loss, depicted in \cref{fig:calc_demo} by the green and purple shading of the subsequent inelastically-scattered wavefields. The intensities from each different final state of each different atom are added incoherently to give the overall core-loss PACBED pattern. Following \cite{PhysRevResearch.1.033186}, the elastic scattering can be described efficiently via scattering matrices. Thus for each incident probe position, the intensity at the detector plane in the far-field may be written
\begin{align}
    I_A(\mathbf{k}_\perp) = \sum_{A', fi} |\mathcal{F}\{ S_2 H^{A'}_{A,fi}S_1\psi_0 \}(\mathbf{k}_\perp)|^2 \;.
    \label{eq:total_intensity}
\end{align}
Note that, to focus on the main ideas, we use a simplified notation here, with $A$ denoting the species of interest and $A'$ denoting the specific atomic sites on which it occurs. It is to be understood that $S_1$ effects elastic scattering of the incident wavefield $\psi_0$ to the depth of the site $A'$ at which the ionization occurs and that $S_2$ effects elastic scattering of the inelastically scattered wavefield to the exit surface. In addition, $\mathcal{F}$ denotes Fourier transform to the diffraction plane, the summation also runs over all (unobserved) initial and final states $fi$ consistent with the energy loss selected, and we have absorbed the interaction constant $\sigma$ into $H^{A'}_{A,fi}$.

Following \citet{coene_inelastic_1990,dwyer_multislice_2005,dalfonso_three-dimensional_2008,DEIMETRY2024114036}, we use a central field, one-electron
wavefunction model for the initial and final states in a spherical harmonic basis and assume transition to continuum states. The bound and unbound atomic electron wavefunction can then be written
\begin{align}
    a_{n\ell m}(r, \theta, \phi) = \frac{P_{n \ell}(r)}{r} Y^m_\ell(\theta, \phi) \nonumber \\
    a_{\varepsilon \ell' m'}(r, \theta, \phi) = \frac{P_{\varepsilon \ell'}(r)}{r} Y^{m'}_{\ell'}(\theta, \phi)
    \label{eq:crystal_wf}
\end{align}
where the bound electron wavefunction $a_{n\ell m}$ is characterised by the principal, orbital and magnetic quantum numbers $n,\ell, m$, the ejected electron wavefunction $a_{\varepsilon \ell' m'}$  is characterised by its kinetic energy $\varepsilon$ and orbital and magnetic quantum numbers $\ell', m'$, $P$ denotes radial wavefunctions and $Y$ denotes spherical harmonics. We use relativistic Hartree-Fock and Hartree-Slater wavefunctions for the bound and continuum radial wave functions respectively \citep{oxley2001atomic}. Selecting a particular ionization edge amounts to choosing $n$ and $\ell$. The particular energy loss above threshold sets $\varepsilon$ (or we could integrate over a range of such energy losses). The summation over final states in \cref{eq:total_intensity} is then over all values of $m$, $\ell'$ and $m'$. While the number of states to be summed is in principle  infinite, the summation converges quickly (\cite{DEIMETRY2024114036} demonstrate the fall-off to be exponential) and so in practice only a limited number of terms are included. Open-source software for simulating core-loss 4D-STEM via this transition potential formulation is available through packages such as \emph{py}$\_$\emph{multislice} \citep{10.1017/S1431927620023326,brown2025py_multislice_repo} and \emph{abTEM} \citep{madsen_abtem_2021,abtem2025abTEM}. Incorporating $S$ matrices as per \citet{PhysRevResearch.1.033186} (implemented in \emph{py}$\_$\emph{multislice}) provides significant speedups. Even so, incorporating elastic scattering after the ionization event for a significant number of inelastic transitions is computationally intensive, and we recommend calculations be done on a CUDA-capable GPU.

\begin{figure}
    \centering
    \includegraphics[width=0.95\linewidth]{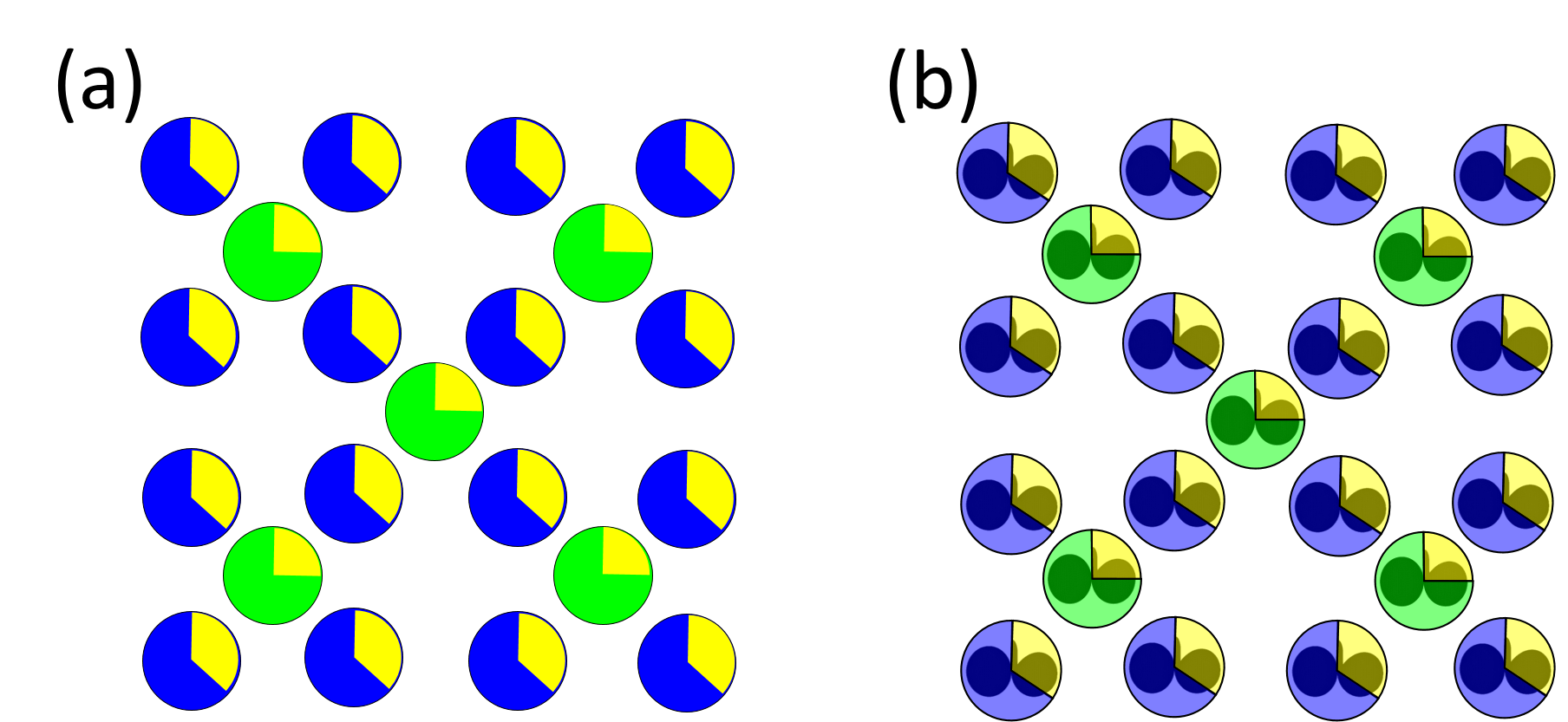}
    \caption{(a) Schematic of a doped generic crystal structure using fractional occupancy within the elastic scattering regime, where each site is partially occupied in proportion to the probability of finding a particular species at that site. (b) Schematic extending the fractional occupancy to the inelastic channels formulation, where the transition potentials, depicted by the spherical harmonic function inset in each of the atoms, are also partitioned based on the probability of finding a particular species at that site.
    }
    \label{fig:frac_occ_view}
\end{figure}

Because handling nonperiodic structures increases the complexity of the calculations, for exploratory simulations of doped crystals it is convenient to invoke the fractional occupancy model, also called the virtual crystal approximation \citep{carlino2005atomic,blom2020probing}, whereby the electrostatic potential of each atomic site is taken to be a sum of the electrostatic potentials of all species that may occupy that site, weighted by the probability of occupation. This is shown schematically in \cref{fig:frac_occ_view}(a), where the proportion of each colour at each atomic site indicates the percentage weighting given to the corresponding atomic species when evaluating the potential at that site. For instance, in the case of a spinel doped with iron the potential at the magnesium sites in the fractional occupancy model is given by 
\begin{align}
    f_{\rm Fe}^{\rm Mg}V_{\rm Fe} + f_{\rm Mg}^{\rm Mg}V_{\rm Mg} \;,
    \label{eq:fracocc_elastic}
\end{align}
where $f_{\rm Fe}^{\rm Mg}$ denotes the fraction of octahedral Mg sites occupied by iron dopants and $f_{\rm Mg}^{\rm Mg}$ denotes the fraction of octahedral Mg sites occupied by Mg atoms after doping. (Throughout this paper, we will use the convention that superscripts denote crystallographic sites and subscripts denote atomic species.) Since only a single dopant species is considered, it follows that $f_{\rm Fe}^{\rm Mg}+f_{\rm Mg}^{\rm Mg}=1$, so these two quantities are not independent. Concentrations can also be calculated from these fractions, for instance the concentration of Fe (in atomic percent) is given by
\begin{align}
c_{\rm Fe} = \frac{N_{\rm Mg} f_{\rm Fe}^{\rm Mg} + N_{\rm Al} f_{\rm Fe}^{\rm Al}}{N} \;,
 \label{eq:concentration_defn}
\end{align}
where $N_{\rm Mg}$, $N_{\rm Al}$ and $N$ respectively denote the total number of Mg, Al and all atoms within a suitably large volume.

The fractional occupancy model applies to the elastic scattering, and so is incorporated when evaluating the scattering matrices $S_1$ and $S_2$ in \cref{eq:total_intensity}. To extend this approach to inelastic scattering we reason as follows. Applying the Born interpretation to \cref{eq:elastic_to_inelastic}, the probability $\text{Pr}_{\rm inel}({\bf r}|B')$ of finding the inelastically scattered electron at ${\bf r}$ in volume $dV$ immediately after ionization of the atom at site $B'$ is given by
\begin{align}
     \text{Pr}_{\rm inel}({\bf r}|B') = \sum_{m\ell'm'} \text{Pr}_{\rm el}({\bf r}) \text{Pr}({\bf r}; n\ell m \rightarrow \varepsilon \ell'm'|B')
\end{align}
where $\text{Pr}_{\rm el}({\bf r})$ is the probability of finding the elastically scattered electron immediately prior to ionization and $\text{Pr}({\bf r}; n\ell m \rightarrow \varepsilon \ell'm'|B')$ is the probability of that specific inelastic transition, related to the transition potential at site $B'$. Fractional occupancy can then be introduced through the conditional probability of the species of interest occupying that site:
\begin{align}
    \text{Pr}( n\ell m \rightarrow \varepsilon \ell'm'|B') &= \sum_A \text{Pr}( n\ell m \rightarrow \varepsilon \ell'm'| B'A) \text{Pr}(BA) \;,
    \label{eq:conditional}
\end{align}
where the sum is over all species $A$ that could occur at site $B'$. But the probability of species $A$ occupying site $B'$ is just the fractional occupancy of dopant species $A$ on the sites of the host species $B$, i.e. $\text{Pr}(BA)=f^B_A$ (the generalisation to handle potentially different fractional occupancy $f^{B'}_A$ at different atomic sites $B'$ is possible, but the additional notation required is cumbersome and so in the interests of clarity is not considered here). Accounting for doping in inelastic scattering thus involves the substitution $H\rightarrow \sqrt{f} H$ (since $\text{Pr}( n\ell m \rightarrow \varepsilon \ell'm')\propto |H|^2$). \Cref{eq:total_intensity} can then be written
\begin{align}
    I_A(\mathbf{k}_\perp) = \sum_{B', fi} f^B_A |\mathcal{F}\{ S_2 H^{B'}_{A,fi}S_1\psi_0 \}(\mathbf{k}_\perp)|^2 \;,
    \label{eq:total_intensity_fracocc}
\end{align}
where $B'$ runs over all sites, and the sum over $A$ in \cref{eq:conditional} drops out by considering the ionization of a single species. This is depicted in \cref{fig:frac_occ_view}(b), where the inset spherical harmonics indicate that the transition potentials are also weighted based on the occupancy. However, we stress that for inelastic scattering it is the resultant intensities that are added, not the potentials, and that (unless there is appreciable overlap of the peaks) we would usually only consider the signal resulting from a single species in any given calculation.

\begin{figure*}
    \centering
    \includegraphics[width=0.75\linewidth]{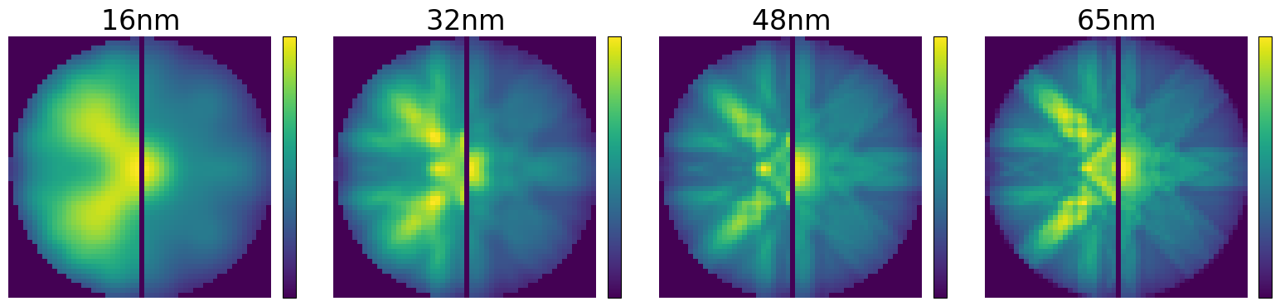}
    \caption{Demonstration of spinel magnesium K ICP contrast with crystal thickness. The crystal is required to be thick enough to ensure channelling which produces stronger contrast. ICPs shown correspond to core-loss PACBED with 30mrad probe forming aperture, 33mrad detector semi-opening angle and 300keV accelerating voltage. Sufficient contrast is achieved at around 32nm.}
    \label{fig:thickness_series}
\end{figure*}

\section{ALCHEMI and (tilt-independent) $k$-factors}\label{sec:tilt_indep_k}

Consider a plane-wave electron beam incident upon a crystal near a low order zone axis orientation. Dynamical diffraction causes the probe electron intensity to increase at some locations and decrease at others, a process called channelling.\footnote{In the context of positive ion scattering, ``channelling'' referred to the ions travelling down open channels between atoms in a material \citep{morgan1973}. The analogy for electrons, which have the opposite charge, is to travel down atomic columns \citep{spence1992channelling}. But in high energy electron microscopy the term is frequently used in a looser sense as being synonymous with dynamical diffraction, irrespective of where the resultant peaking up of electron intensity occurs \citep{spence1992channelling,buseck1992electron}.} As per \cref{eq:total_intensity_fracocc}, the strength of the inelastic signal depends on the overlap between the transition potentials and the probe electron distribution. Varying the orientation of the incident beam changes the channelling conditions leading to a ``channelling enhancement'' of the inelastic signal. ALCHEMI exploits this variation to determine the site occupancy of dopant species through reference to the variation in signal of the species at the potential host sites in the crystal.

In the original ALCHEMI method, comparing X-ray signals from a weakly channelling condition to those from a strongly channelling condition was used to deduce what fraction of some dopant was present at each of two possible sites. This method used measured signals alone, did not require the orientations used to be precisely known 
and involved no unknown or adjustable parameters, but it did make some assumptions, most notably that the inelastic interactions were highly localised \citep{spence1988localization}. The statistical ALCHEMI approach of Rossouw and co-workers (see, e.g., \cite{rossouw_statistical_alchemi,rossouw1996statistical,ofori2005determining}) 
used measurements across a range of tilts to improve the statistical robustness and generalise readily to multiple dopant species and host sites, and to determine absolute concentrations at the expense of needing to know the Cliff-Lorimer $k$-factors \citep{cliff1975quantitative} that relate the ratio of inelastic signals to the ratio of concentrations in the absence of channelling (which, in the case of X-rays, depend not only on the ionization cross-sections but also on fluorescence yields and microscope-dependent detector efficiencies at the pertinent energies). Neither method relies on crystal thickness being known, though it must be large enough that appreciable dynamical scattering occurs. As a concrete example, \cref{fig:thickness_series} shows simulated Mg-K shell ICPs in our spinel test case which display increasing structural detail with increasing crystal thickness due to dynamical diffraction.

To aid later discussion, we present a simplified derivation of the key equation for statistical ALCHEMI assuming a single dopant species $X$ that can occupy only two possible crystallographic sites in the host lattice, $A$ and $B$. The inelastic signal $I_X$ can (conceptually) be decomposed into the contributions from the atoms $X$ at $A$ sites, denoted $I^A_X$, and at $B$ sites, denoted $I^B_X$, as
\begin{equation}
    I_{X} = I^A_X + I^B_X \;.
    \label{eq:ALCHEMI_deriv00}
\end{equation}
As a specific example of this notation, $I_{\rm Fe}^{\rm Mg}$ denotes the intensity from Fe atoms occupying octahedral sites which are nominally occupied by Mg. With the reasons for writing it this way to become clearer presently, it follows trivially from \cref{eq:ALCHEMI_deriv00} that
\begin{equation}
    I_{X} = \frac{I^A_X}{I_A}I_A + \frac{I^B_X}{I_B}I_B \;.
    \label{eq:ALCHEMI_deriv01}
\end{equation}
In the absence of channelling, the inelastic intensity at each site is proportional to the concentration of the element at that site, e.g. $I^A_X \propto f^A_X$, where, as described in conjunction with \cref{eq:fracocc_elastic}, $f^A_X$ denotes the fraction of sites $A$ occupied by dopant $X$. The constant of proportionality depends on factors such as the cross-sections and the detector geometry. When comparing intensities and concentrations of different species, the constants of proportionality can be gathered to form the Cliff-Lorimer $k$-factor \citep{cliff1975quantitative,willans_carter}
\begin{align}
    \frac{f_A^A}{f_X^A} = k_{AX} \frac{I_A}{I_X^A} \;.
    \label{eq:k_factor}
\end{align}
(The Cliff-Lorimer $k$-factor is usually linked to a ratio of concentrations in weight percent, but it is recognised that other choices are sometimes more convenient \citep{mellini1985proportionality,willans_carter} and here we opt to work with fractional site occupancy, as per \cref{eq:fracocc_elastic}.)

Using \cref{eq:k_factor}, \cref{eq:ALCHEMI_deriv01} can be written
\begin{align}
    I_{X} = \frac{f_X^A}{f_A^A} k_{AX} I_A + \frac{f_X^B}{f_B^B} k_{BX} I_B \;.
    \label{eq:ALCHEMI_deriv02}
\end{align}
The notation of \cref{eq:ALCHEMI_deriv02} suggests a single intensity for each species, but in the experimental geometries of \cref{fig:ICPs_showcase} the ICPs are functions of orientation in the CTEM case or detector pixel coordinate in the STEM case. Assuming (questionably, as we will later show) we can introduce this angle dependence into the ICPs without introducing it into the $k$-factors, \cref{eq:ALCHEMI_deriv02} becomes:
\begin{align}
    I_{X}(\vec \theta) = \frac{f_X^A}{f_A^A} k_{AX} I_A(\vec \theta) + \frac{f_X^B}{f_B^B} k_{BX} I_B(\vec \theta) \;.
    \label{eq:STAT_ALCHEMI}
\end{align}

This equation predicts that the ICP for the dopant, $I_{X}(\vec \theta)$, will be a linear combination of the ICPs for species on the potential dopant sites in the host lattice, $I_A(\vec \theta)$ and $I_B(\vec \theta)$. Moreover, we can solve an over-constrained (by data from all the different $\vec \theta$ values) system of simultaneous linear equations (for instance via $\chi^2$ minimization) to obtain a set of concentrations $\{f\}$, provided the $k$-factors are known. Early literature with plane-wave incidence sought to determine these $k$-factors via an off-axis experiment, and experimental determination is useful for EDX ALCHEMI because there are a number of factors that may not be accurately known, including fluorescence yields and detector efficiencies. However, for EELS we can relate the $k$-factors directly to the transition potentials used in our simulations. Assuming plane-wave incidence (as per traditional ALCHEMI), we can take the thickness tending to zero limit\footnote{The conventional construction of the $k$-factors neglects channelling, and thus their values are assumed to be independent of sample thickness.} in \cref{eq:total_intensity_fracocc} by replacing the scattering matrices with identity operators. The overall intensity due to the ionization of reference species $A$ is then
\begin{align}
    I_A(\mathbf{k}_\perp) &= f^A_A \sum_{A', fi} \int |\mathcal{F}\{ H^{A'}_{A,fi}\}(\mathbf{k}_\perp)|^2 d{\bf k} \nonumber \\ &= f^A_A N_A \sum_{fi} \int |H_{A,fi}(\mathbf{r}_\perp)|^2 d{\bf r}_\perp
        \;,
    \label{eq:total_intensity_single_atom}
\end{align}
where the second equality follows from Parseval's theorem and recognising that, with $A'$ restricted to a single plane, the sum over identical terms reduces to a product of one such term and the number of such sites $N_A$.  

Using \cref{eq:k_factor}, the EELS $k$-factor is given by\footnote{The EELS $k$-factors are even more directly related to inelastic ionization cross-sections than those for EDX. As derived by \cite{findlay2008modeling}, the inelastic scattering cross-section at depth $z$ over range $dz$ is given by
\begin{align}
    \sigma \propto \int d\mathbf{r}_\perp \:  |H_{fi}(\mathbf{r}_\perp)|^2
\end{align}
where single (cf. multiple) inelastic scattering is assumed.}
\begin{align}
     k_{AX} = \frac{ \sum_{fi} \int |H_{X,fi}|^2 d\bf \bf r_\perp}{ \sum_{fi} \int |H_{A,fi}|^2 d\bf \bf r_\perp} \;.
     \label{eq:tilt_indep_k_fac}
\end{align}

It is instructive to substitute \cref{eq:tilt_indep_k_fac} into \cref{eq:STAT_ALCHEMI} and divide by $\sum_{fi} \int |H_{X,fi}|^2 d\bf \bf r_\perp$, yielding
\begin{align}
     &\frac{I_{X}(\vec \theta)}{\sum_{fi} \int |H_{X,fi}|^2 d\bf \bf r_\perp} \nonumber\\
     &= \frac{f_X^A}{f_A^A} \frac{I_A(\vec \theta)}{\sum_{fi} \int |H_{A,fi}|^2 d\bf \bf r_\perp} + \frac{f_X^B}{f_B^B} \frac{I_B(\vec \theta)}{\sum_{fi} \int |H_{B,fi}|^2 d\bf \bf r_\perp} \;.
    \label{eq:ALCHEMI_normalised}
\end{align}
This form highlights the role of the $k$-factor as being to ``correct'' for the difference in scales between the different elements' cross-sections, while preserving the pattern that arises from the variation in $\vec{\theta}$.

Using \cref{eq:STAT_ALCHEMI,eq:tilt_indep_k_fac} together with the ICPs in \cref{fig:ICPs_showcase}(b), (e) and (h), $\chi^2$ minimization accurately retrieves the concentrations assumed in those simulations. However, it fails when used on  the ICPs in \cref{fig:ICPs_showcase}(c) and (f). 

\subsection{Break-down of (tilt-independent) $k$-factors}
That the $k$-factors in \cref{eq:tilt_indep_k_fac} fail in certain circumstances is expected with careful consideration of the inelastic process. The nature of ionization is inherently tied to the characteristics of the transition potentials: their spatial extent governs the interaction range, while their amplitude dictates the probability of excitation. As per \cref{eq:ALCHEMI_normalised}, the conventional $k$-factors account for the latter. However, they only account for the former under limited circumstances.

One such circumstance is if the interaction potentials are highly localised. To see this, consider substituting $H^{A'}_{X,fi}({\bf r}_\perp)=h_{X,fi} \delta({\bf r}_\perp-{\bf r}^{A'}_\perp)$ into \cref{eq:total_intensity_fracocc}:
\begin{align}
    I^A_X(\mathbf{k}_\perp) = f^A_X \left[\sum_{fi}|h^X_{fi}|^2\right] \sum_{A'} |\mathcal{F}\{ S_2 \delta({\bf r}_\perp-{\bf r}^{A'}_\perp) S_1\psi_0 \}(\mathbf{k}_\perp)|^2 \;.
    \label{eq:deltafn}
\end{align}
The orientation (${\bf k}_\perp$) dependence comes entirely through the factor $|\mathcal{F}\{ S_2 \delta({\bf r}_\perp-{\bf r}^{A'}_\perp) S_1\psi_0 \}(\mathbf{k}_\perp)|^2$, which in this limit is independent of the species $X$ present at sites $A$. Thus the pattern from all elements at $A$ sites, including that from species $A$, differ only in amplitude and not in structure. This orientation dependence then cancels in taking the intensity ratio in the $k$-factor definition of \cref{eq:k_factor}, thus keeping that equation valid. This was recognised in the early ALCHEMI literature and was the basis of favouring EDX signals and/or those from tightly bound shells, for which assuming a highly localised interaction tends to be a good approximation \citep{spence1988localization,buseck1992electron}.

Rossouw and coworkers recognised that this would also be true if all interaction potentials of interest had identical widths \citep{rossouw_statistical_alchemi,rossouw1996statistical,ofori2005determining}. To see this, consider substituting $H^{A'}_{X,fi}({\bf r}_\perp)= h_{X,fi} \overline{h}({\bf r}_\perp-{\bf r}^{A'}_\perp)$ into \cref{eq:total_intensity_fracocc} restricted to a single site $A$:
\begin{align}
    I^A_X(\mathbf{k}_\perp) = f^A_X \left[\sum_{fi}|h_{X,fi}|^2\right] \sum_{A'} |\mathcal{F}\{ S_2 \overline{h}({\bf r}_\perp-{\bf r}^{A'}_\perp) S_1\psi_0 \}(\mathbf{k}_\perp)|^2 \;.
    \label{eq:samewidth}
\end{align}
Similar to \cref{eq:deltafn}, the orientation (${\bf k}_\perp$) dependence now comes entirely through the factor $|\mathcal{F}\{ S_2 \overline{h}({\bf r}_\perp-{\bf r}^A_\perp) S_1\psi_0 \}(\mathbf{k}_\perp)|^2$, which is independent of the species $X$ present at sites $A$ by the assumption that the distribution $\overline{h}$ is common to all interaction potentials of interest. Thus the patterns from all elements at $A$ sites again differ only in amplitude and not in structure, and the $k$-factor definition is again valid.

However, for EELS in realistic systems the interaction ranges differ across the various constituent elements. As a result, the ICPs differ not only in amplitude but also in structure, and thus \cref{eq:STAT_ALCHEMI} no longer holds: $I_{X}(\vec \theta)$ ceases to be a linear combination of the ICPs for species on the potential dopant sites in the host lattice, $I_A(\vec \theta)$ and $I_B(\vec \theta)$. \citet{rossouw1996statistical,OXLEY1999109} show some examples of the errors in statistical ALCHEMI resulting from the difference in interaction ranges between elements. Let us here further explore this issue in the context of our particular case study.

We can quantify the sensitivity and breakdown of a linear model by modifying the Fe-doped MgAl$_2$O$_4$ test case by replacing the transition potentials with Gaussians whose relative widths (parameterised by the standard deviation $\sigma$ of the Gaussian) we will allow to vary. Since the transition potentials of neighboring atoms on the periodic table are very similar, we reduce the parameter space by setting the widths of the Gaussians on the Mg and Al sites to be identical, $\sigma_{\rm Mg}=\sigma_{\rm Al}$. On the assumption of comparing equivalent shells, we will also consider the case $\sigma_{\rm Mg/Al} > \sigma_{\rm Fe}$, since equivalent shells in atoms with higher atomic number will have higher ionization threshold and therefore be more localised.

\begin{figure}
    \centering
    \includegraphics[width=1\linewidth]{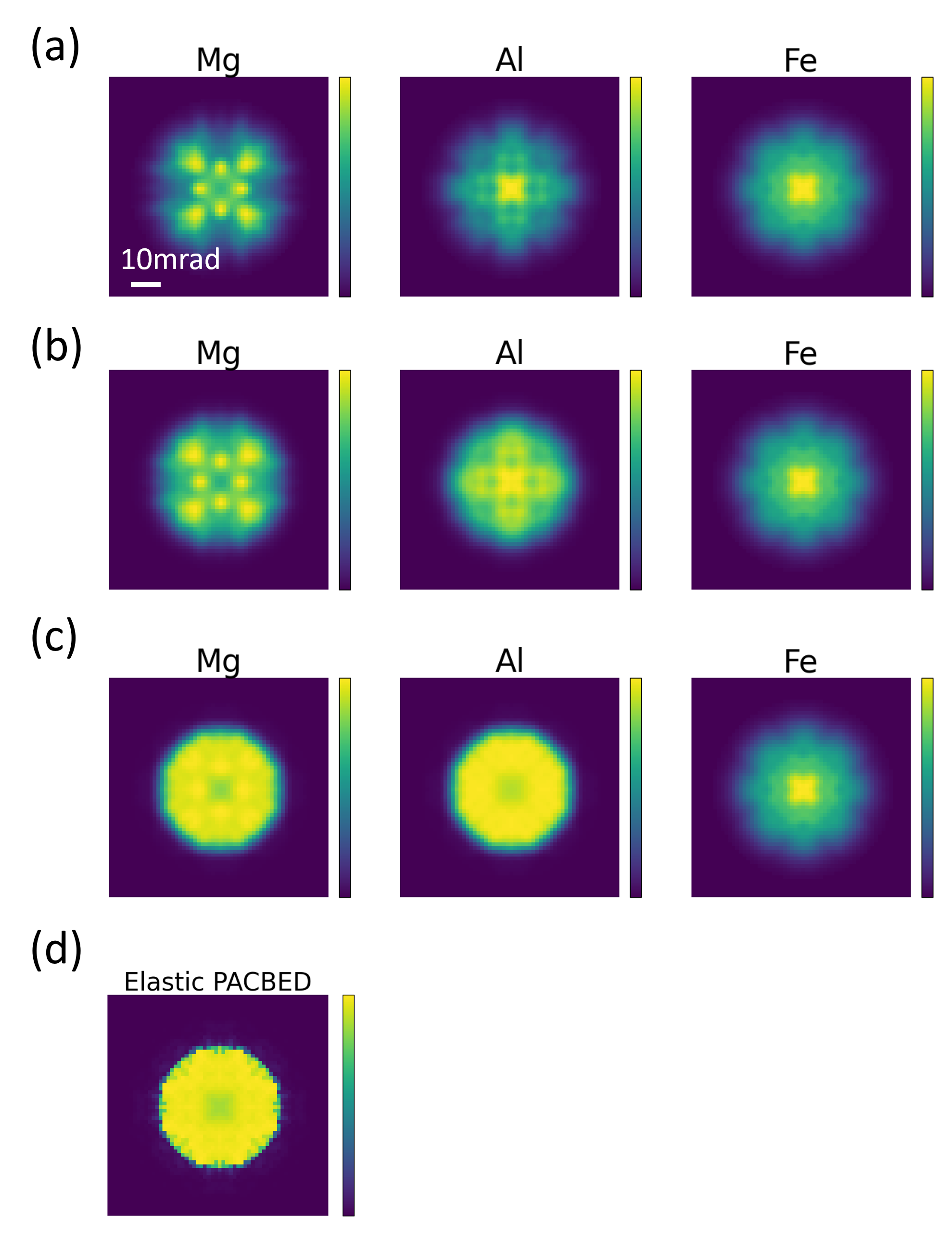}
    \caption{ICPs of an iron-doped spinel with 5\% Fe doping on magnesium sites and 7\% on aluminium sites, where the transition potentials were replaced with Gaussians characterized by their standard deviations, where  $\sigma_{\rm Fe}=0.5${\AA} and $\Delta \sigma=\sigma_{\rm Mg/Al} - \sigma_{\rm Fe} \geq 0${\AA}. A 20mrad probe forming aperture and a 30mrad detector are assumed. (a) $\Delta\sigma=0.0${\AA}, corresponding to an exact linear model with unique solution corresponding to the correct doping concentration. (b) $\Delta\sigma=0.2${\AA}, corresponding to a failing linear model with unique least-squares solution. (c) $\Delta\sigma=0.8${\AA}, corresponding to a failed linear model with many least-squares solutions. Observe that the core-loss PACBEDs increasingly resemble the (d) elastic PACBED with increasing delocalization.}
    \label{fig:ICP_gaussian_eg}
\end{figure}

\Cref{fig:ICP_gaussian_eg} shows ICPs simulated with Gaussian transition potentials. We set $\sigma_{\rm Fe}=$ 0.5{\AA} for all cases, and successive rows consider increasingly large values of $\Delta\sigma=\sigma_{\rm Mg/Al} - \sigma_{\rm Fe}$: $\Delta\sigma=0.0${\AA} in \cref{fig:ICP_gaussian_eg}(a), $\Delta\sigma=0.2${\AA} in \cref{fig:ICP_gaussian_eg}(b) and $\Delta\sigma=0.8${\AA} in \cref{fig:ICP_gaussian_eg}(c). In \cref{fig:ICP_gaussian_eg}(a), the interaction ranges are identical and the Fe ICP is an exact linear combination of the Mg and Al ICPs. In \cref{fig:ICP_gaussian_eg}(b), a cross-shaped feature appears in the Al ICP but no compensating feature appears in the Mg ICP. The Fe ICP thus ceases to be an exact linear combination of the Mg and Al ICPs, though, as we quantify below, an approximate linear combination may still give reasonable concentration estimates. In \cref{fig:ICP_gaussian_eg}(c), the contrast of the Mg and Al ICPs is almost reversed from that of the Fe ICP --- the linear model has broken down completely and concentration estimation is impossible. (It is interesting to note that for the wide Gaussians assumed for Mg and Al in \cref{fig:ICP_gaussian_eg}(c), the ICPs closely resemble the elastic PACBED shown in \cref{fig:ICP_gaussian_eg}(d). The resemblance would be exact if the transition potentials in \cref{eq:elastic_to_inelastic} were unity, i.e. completely delocalised.) 

\begin{figure}
    \centering
    \includegraphics[width=1\linewidth]{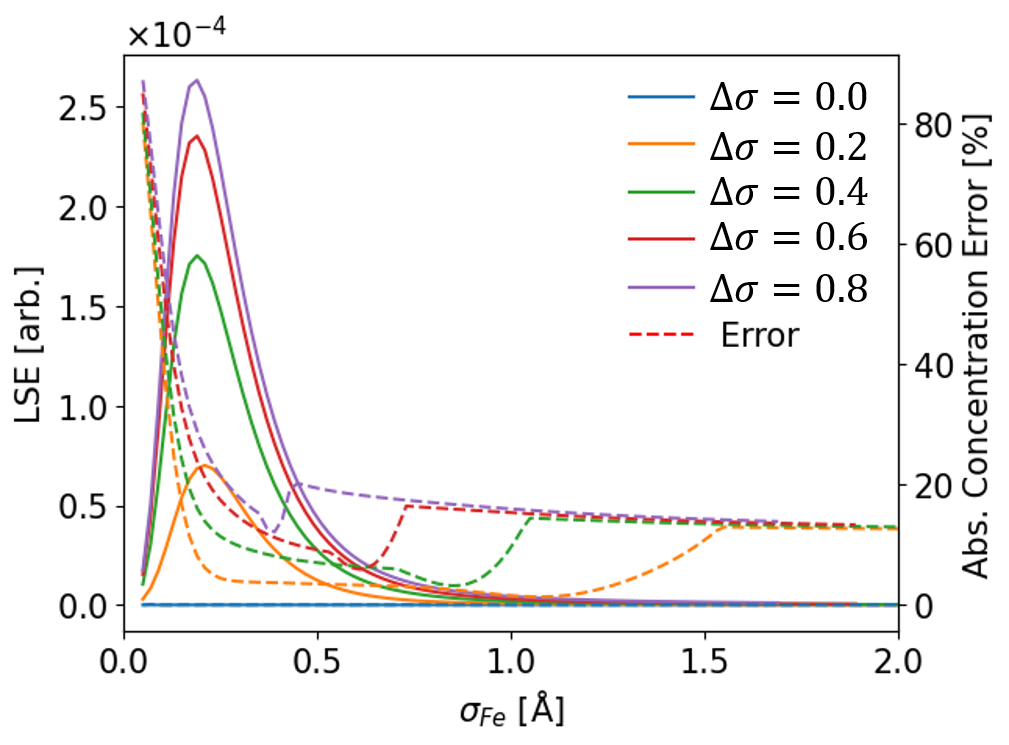}
    \caption{Least-squares error (LSE, solid lines) with positive fitting coefficients constraint and absolute dopant concentration percentage error (dashed lines) as determined using \cref{eq:STAT_ALCHEMI} applied to various triplets of ICPs of an iron-doped spinel with 5\% on magnesium sites and 7\% on aluminium sites, assuming a 20mrad probe forming aperture and 30mrad detector. The transition potentials are replaced with Gaussians characterized by their standard deviations $\sigma$, where $\Delta \sigma=\sigma_{\rm Mg/Al} - \sigma_{\rm Fe}$. Three regions of behavior are observed. Within the first region, the linear model holds fairly well (relatively low LSE) but the absolute dopant concentration percentage error is nevertheless very large. In the second region, the absolute dopant concentration percentage error is lower, despite the linear model holding less well (relatively large LSE). In the third region, the linear model now holds fairly well again (relatively low LSE), but the problem becomes ill-posed and the absolute dopant concentration determination unreliable.}
    \label{fig:lse}
\end{figure}

To summarise the breakdown of the linear model and the reliable estimation of concentrations, \cref{fig:lse} shows the least-squares error (LSE) of fitting \cref{eq:STAT_ALCHEMI} (solid lines) and the absolute percentage error in the deduced dopant concentration (dashed lines) as a function of the standard deviation of the dopant's Gaussian potential, $\sigma_{\rm Fe}$, for various values of $\Delta\sigma=\sigma_{\rm Mg/Al} - \sigma_{\rm Fe}$. As anticipated in the idealized scenario where the interaction ranges are identical, i.e. $\Delta \sigma = 0${\AA}, the linear combination is exact and the concentration accurately determined. For $\Delta \sigma > 0${\AA}, the graph can be divided into three broad regions of the behaviour of the absolute concentration error across $\sigma_{\rm Fe}$. In the first region, $\sigma_{\rm Fe} \lesssim 0.5${\AA}, the absolute concentration error is largest. In the second region, the absolute concentration error levels off around a minimum, and in the third it rises sharply then plateaus. 

The absolute concentration error can be calculated in this exploration because the true result is known, but this will not be the case in practice. The LSE can be calculated in practice, but unfortunately \cref{fig:lse} shows that the LSE does not correlate with the absolute concentration error. In the limit of large $\sigma_{\rm Fe}$, the LSE tends to zero because with increasingly wide transition potentials the contrast in all ICPs tends to that of the elastic PACBED pattern: the linear problem becomes ill-posed, and the least squares solution is unlikely to correspond to the true concentration.

\begin{figure}
    \centering
    \includegraphics[width=1\linewidth]{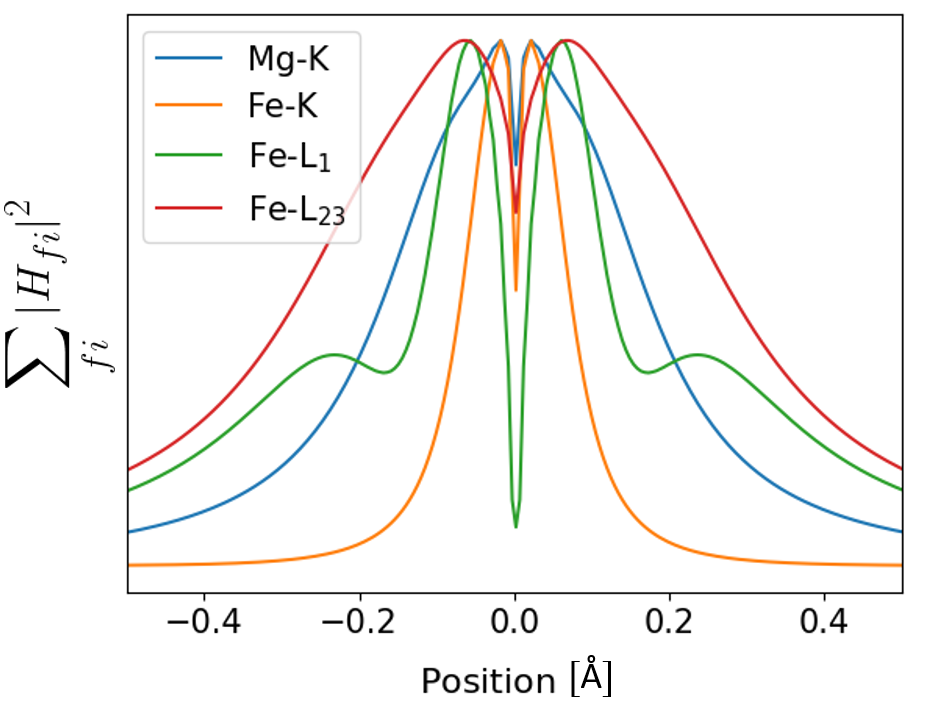}
    \caption{Line profiles through the sum of the mod-squares of the relevant transitions potentials for the Mg-K shell and the Fe-K, Fe-L$_1$ and Fe-L$_{2,3}$ shells. The maximum amplitudes are normalized so the widths can be compared. The Fe shell with width closest to that of the Mg-K shell is the Fe-L$_1$ shell.}
    \label{fig:transition_potentials}
\end{figure}

Having explored the breakdown in EELS ALCHEMI using Gaussians, we proceed to using accurate transition potentials. \Cref{fig:transition_potentials} shows the sum of the square-modulus of the transition potentials for the Mg-K, Fe-K, Fe-L$_1$ and Fe-L$_{2,3}$ shells. The figure uses the same value of the energy above threshold $\varepsilon$ for all transitions, the choice of which is innocuous in comparison to the core-loss ionization threshold and therefore has a minimal effect on the potentials.\footnote{For sufficiently low atomic numbers, where typically accessible energies above threshold $\varepsilon$ are a larger fraction of the threshold energy, $\varepsilon$ may have more influence. In such cases, setting $\varepsilon$ as large as possible may help a little in localising the transition potentials.} The Fe shell with width closest to that of the Mg-K shell is the L$_1$ shell, followed by the L$_{2,3}$ shell. Since EELS ALCHEMI accuracy is sensitive to differences in width between the transition potentials, we would predict that using the Fe-L$_1$ ICP in conjunction with the Mg-K and Al-K ICPs should produce more accurate concentrations than using either the Fe-K or Fe-L$_{2,3}$ ICPs. 

\begin{figure*}
    \centering
    \includegraphics[width=1\linewidth]{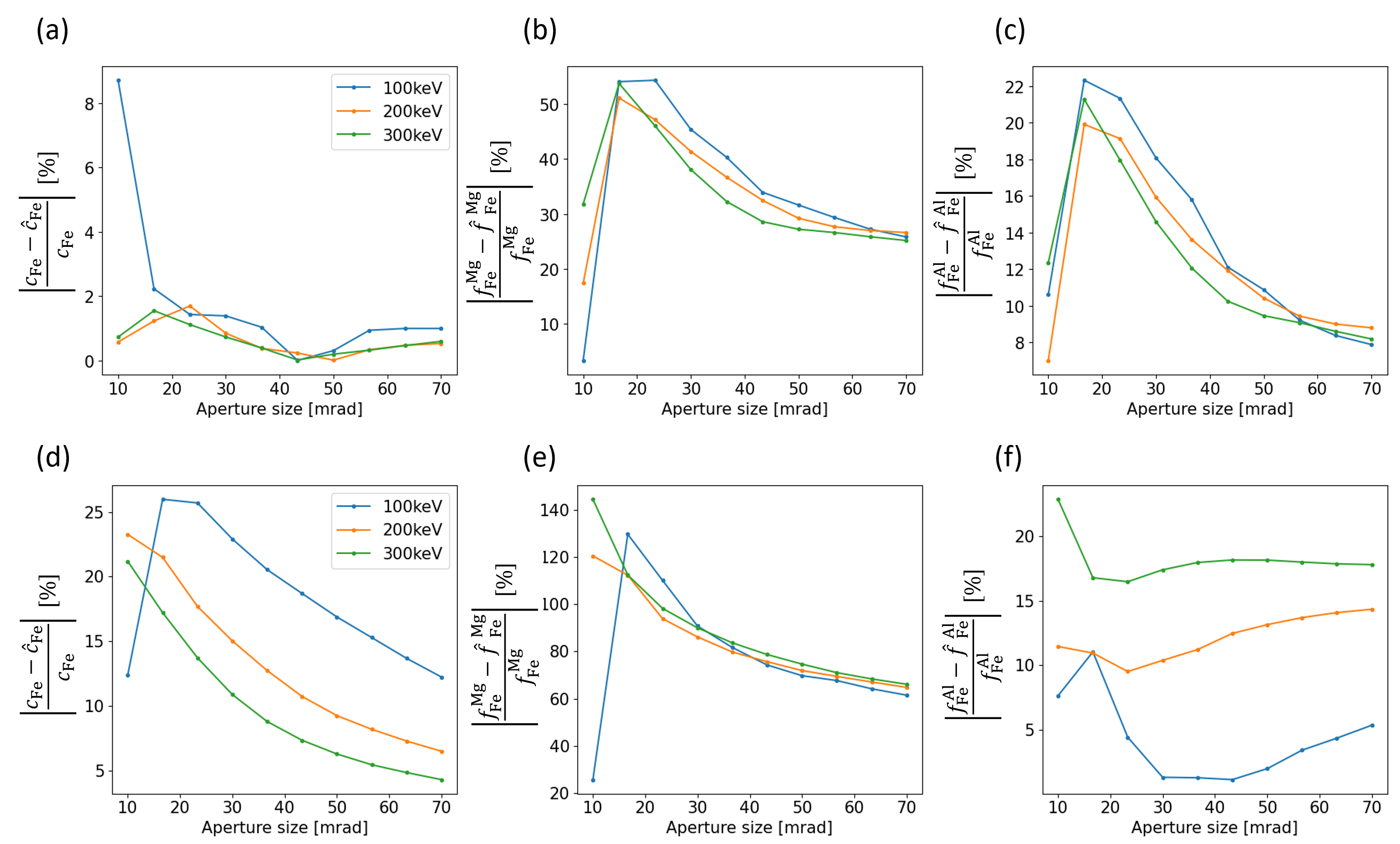}
    \caption{Absolute dopant concentration percentage errors for (a) all sites, (b) Mg sites and (c) Al sites applying statistical EELS ALCHEMI to simulated (infinite dose) core-loss PACBED ICPs for the Mg-K, Al-K and Fe-L$_1$ shell from a spinel doped with 5\% Fe on Mg sites and 7\% Fe on Al sites. Quantities with carets denote least-squares solutions. The detector is chosen to be 10\% larger than the aperture size. All calculations assume $\varepsilon=10$eV. 
    (d)-(f) As per (a)-(c), but for Fe-L$_{2,3}$ instead of Fe-L$_1$. While the L$_1$ shell (a-c) is predicted to produce more accurate concentrations than using the L$_{2,3}$ shell (d-f), it is not accessible experimentally.
    }
    \label{fig:hn0_ALCHEMI_variation}
\end{figure*}

Since the integrated intensity is an interplay between the transition potentials and the surrounding current density  which relates to the incident wavefield, parameters that influence these quantities will influence the accuracy of EELS ALCHEMI. Specifically, the beam energy influences both the incident wavefield and transition potentials, while the probe-forming aperture only influences the incident wavefield. \Cref{fig:hn0_ALCHEMI_variation} shows the absolute percentage error in the dopant concentration as a function of probe forming aperture and accelerating voltage, determined from simulated ICPs for the Mg-K and Al-K shells and the Fe-L$_1$ shell (a)-(c) and Fe-L$_{2,3}$ shell (d)-(f). Transitions up to $\ell'=2$ (inclusive) were simulated for Mg-K, Al-K and Fe-L$_1$. For computational feasibility, we have opted to only include select transitions of the Fe-L$_{2,3}$ shell which contribute more than 1\% to the overall signal. These transitions are $(m,\ell',m')=(-1,1,-1)$, $(1,1,1)$, $(-1,2,-2)$, $(-1,2,-1)$, $(0,2,-1)$, $(-1,2,0)$, $(0,2,0)$, $(1,2,0)$, $(0,2,1)$, $(1,2,1)$, $(1,2,2)$. These same transitions are used for all subsequent calculations for these shells unless stated otherwise.

For the Fe-L$_1$ edge, \cref{fig:hn0_ALCHEMI_variation}(a)-(c), maximizing the aperture size and accelerating voltage improves accuracy. This is expected, since increasing the aperture size sharpens the probe, reducing the influence of the interaction range on the measured intensity since a narrow probe will need to be near the atomic site to induce core-loss scattering. Increasing beam energy has the contrary effect of smoothing the probe, and so the potentials look more delta function-like relative to the scale of the probe.

The overall concentration error shown in \cref{fig:hn0_ALCHEMI_variation}(a) is smaller, as a percentage, than the concentration error on individual sites shown in \cref{fig:hn0_ALCHEMI_variation}(b) and (c). If the dopant ICP were a linear combination of the reference site ICPs, which may not be strictly true here, $c_{\rm Fe}$ being the most robust can be understood by considering the statistical ALCHEMI prediction of the corresponding uncertainty \citep{rossouw_statistical_alchemi} in \cref{eq:concentration_defn} as
\begin{align}
    \delta c_{\rm Fe}^2 = \frac{N^2_{\rm Mg}}{N^2} \left(\delta f_{\rm Fe}^{\rm Mg} \right)^2 + \frac{N^2_{\rm Al}}{N^2} \left(\delta f_{\rm Fe}^{\rm Al} \right)^2
    \label{eq:quadrature}
\end{align}

where $\delta f_{\rm Fe}^{\rm Mg}$ is the uncertainty in $f_{\rm Fe}^{\rm Mg}$ and likewise for Al.\footnote{When the linear combination breaks down, the uncertainties will not be independent and so strictly \cref{eq:quadrature} should be amended to include a covariance term, but the number-of-atoms dependence remains similar and so the qualitative conclusion remains reasonable.} Since the errors are added in quadrature and weighted by the concentration of the corresponding reference species, the overall concentration is expected to have the smallest uncertainty of all solved quantities. 

For the Fe-L$_{2,3}$ edge, \cref{fig:hn0_ALCHEMI_variation}(d)-(f), the errors are substantially larger, and no clear trends with aperture size or accelerating voltage are evident. This is less expected, but we attribute it to the width of the Fe-L$_{2,3}$ transition potentials being substantially different to those of Mg-K and Al-K (see \cref{fig:transition_potentials}), with the consequence that the linear model is a poor approximation.

When the difference in the transition potential widths is sufficiently large that the ICP for the dopant cannot reliably be assumed to be a linear combination of the ICPs of the host species then an alternative approach is needed. Our test case of iron-doped spinel, for 300keV electrons and a 30mrad probe-forming aperture, is such a case since the Fe-L$_1$ shell is inaccessible experimentally (its cross-section is too low) and \cref{fig:hn0_ALCHEMI_variation}(d)-(f) shows the errors of applying the existing approach to the Fe-L$_{2,3}$ shell to be large. However, since the structure of the host material is known, we can obtain better results by using simulations as a reference.

\section{Tilt-dependent $k$-factors}\label{sec:tilt_dep_k}

The case study of the previous section shows a general result: when the localization of the different shells (i.e. the widths of the transition potentials) differs, the traditional $k$-factor and  statistical ALCHEMI approaches break down. Various corrections have been proposed in the EDX ALCHEMI literature to overcome this \citep{PENNYCOOK1988239,walls1992formulation,anderson1997alchemi}. In particular, \citet{OXLEY1999109} compare corrections at four different levels of approximation, the most general of which relies on detailed simulations for reference. In this section we adapt that most general approach for EELS ALCHEMI using core-loss PACBED patterns. To this end, we revisit the ALCHEMI derivation of \cref{eq:ALCHEMI_deriv00,eq:STAT_ALCHEMI}, but now assume the orientation dependence of the ICP intensities at the outset.

\begin{figure*}
    \centering
    \includegraphics[width=1\linewidth]{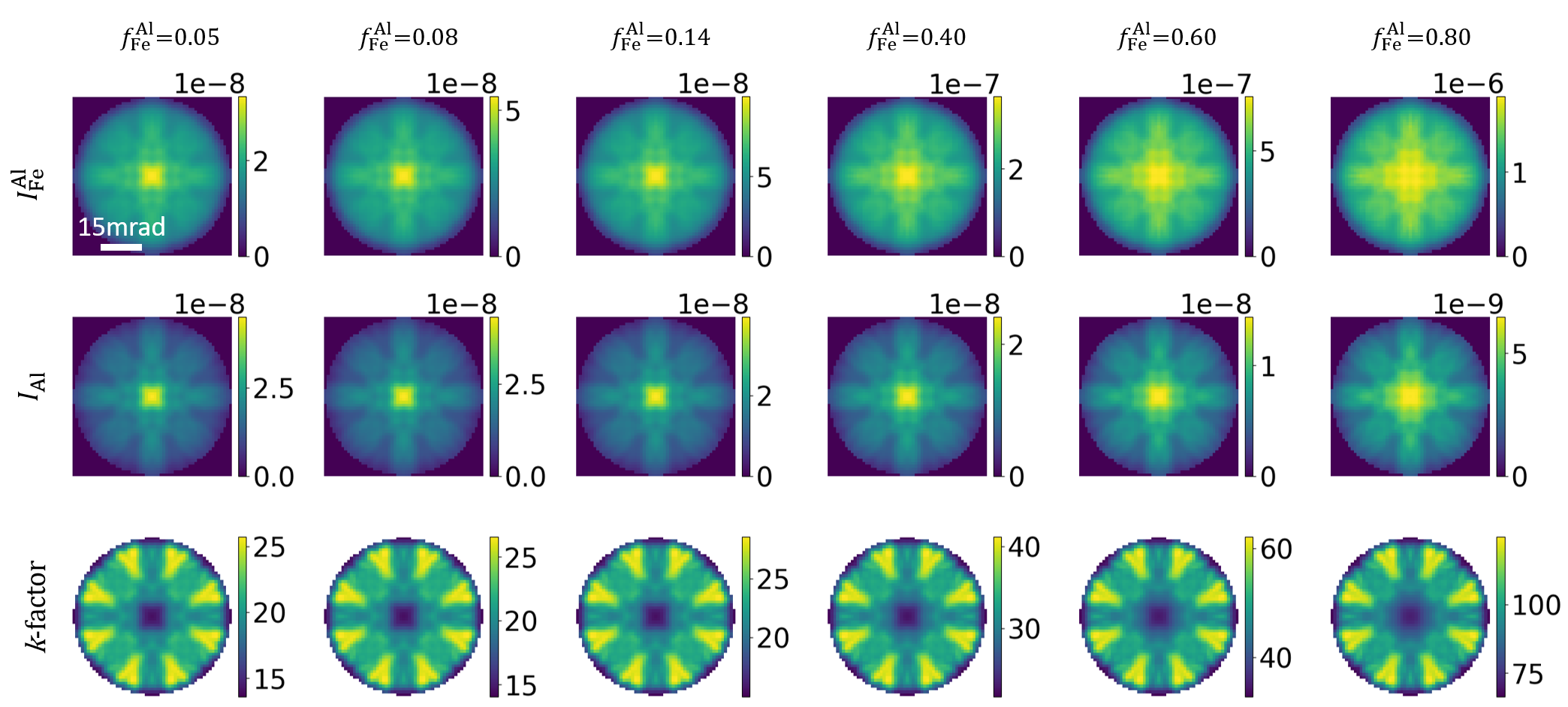}
    \caption{Variation in tilt-dependent Fe-doped spinel $k$-factors with increasing Fe dopant concentration on Al sites along the columns while the dopant on Mg columns is held fixed at $f_{\rm Fe}^{\rm Mg}=0.05$. The first two rows correspond to the Al-K and Fe-L$_{2,3}$ ICPs respectively. The third column is the corresponding $k$-factor. Observe that for small concentrations the ICPs are not perceptibly different. For large concentrations, the individual ICPs change perceptibly, which is seen to change the scale of the corresponding $k$-factors somewhat, though their contrast (structural detail) differs little from those at lower concentrations. All calculations assume a 30mrad probe forming aperture, 300keV electrons and $\varepsilon=10$eV.} 
    \label{fig:k_fac_v_conc}
\end{figure*}

In the case of the spinel doped with iron,
\begin{align}
    I_{\rm Fe}(\vec \theta) &= I_{\rm Fe}^{\rm Mg}(\vec \theta) + I_{\rm Fe}^{\rm Al}(\vec \theta) \\
     & = \frac{I_{\rm Fe}^{\rm Mg}(\vec \theta)}{I_{\rm Mg}(\vec \theta)} I_{\rm Mg}(\vec \theta)+ \frac{I_{\rm Fe}^{\rm Al}(\vec \theta)}{I_{\rm Al}(\vec \theta)} I_{\rm Al}(\vec \theta) \;.
\end{align}
Unlike the earlier derivation, the fractional terms now depend on the dopant concentration in two ways. As before, the weight of the transition potentials at each site are reduced based on the concentration --- this leads to the prefactor $f^B_A$ (the fractional occupancy of dopant species $A$ on site $B$) in \cref{eq:total_intensity_fracocc}. In addition, the presence of dopants along the column further modifies the elastic scattering before and after each inelastic event --- this is encoded in the scattering matrices in \cref{eq:total_intensity_fracocc}. Normalising out the former by analogy with the earlier derivation, we obtain

\begin{align}
    I_{\rm Fe}(\vec \theta)= \frac{f_{\rm Fe}^{\rm Mg}}{f_{\rm Mg}^{\rm Mg}} k_{\rm Mg, Fe}(\vec \theta) I_{\rm Mg}(\vec \theta)+ \frac{f_{\rm Fe}^{\rm Al}}{f_{\rm Al}^{\rm Al}} k_{\rm Al, Fe}(\vec \theta) I_{\rm Al}(\vec \theta) \;,
    \label{eq:general_ALCHEMI_spinel}
\end{align}
where we have now defined \emph{tilt-dependent} $k$-factors by
\begin{align}
    \frac{f_A^A}{f_X^A} = k_{AX}(\vec \theta) \frac{I_A(\vec \theta)}{I_X^A(\vec \theta)}
    \label{eq:k_factor_general}
\end{align}
(cf. \cref{eq:k_factor}).\footnote{For $m$ dopants and $n$ reference sites, \cref{eq:general_ALCHEMI_spinel,eq:k_factor_general} generalize to 
\begin{align}
    I_t(\vec \theta) = \sum_{i=1}^n \frac{f_t^i}{1-\sum_{j=1}^m f_j^i} k_{it}(\vec \theta) I_i(\vec \theta), \quad \frac{f_i^i}{f_t^i} = k_{it}(\vec \theta) \frac{I_i(\vec \theta)}{I_t^i(\vec \theta)}
    \label{eq:general_alchemi_expre}
\end{align}
where $f_t^i$ denotes the fractional occupancy of dopant $t$ on crystallographic reference site $i$, and $t\in \{1,...,m\}$.} \Cref{fig:k_fac_v_conc} shows several tilt-dependent $k$-factors as a function of concentration, and the ICPs from which they were calculated as per \cref{eq:k_factor_general}. The tilt-dependence in the $k$-factors is clearly evident, underscoring why assuming tilt-independent $k$-factors, i.e. approximating the images in the bottom row of \cref{fig:k_fac_v_conc} as having uniform intensity, is a poor approximation here.

Having factored out the fractional occupancy, the remaining concentration dependence comes from the effect of concentration on the elastic channelling. The ICPs in \cref{fig:k_fac_v_conc} show very little variation in contrast (structural detail) with dopant concentration, but the scale grows with the increasing proportion of iron atoms along the column not only because of the increase in the number of iron atoms but also because the higher atomic number of iron relative to aluminium increases the strength of channelling. Consequently, the tilt-dependent $k$-factors show little variation in contrast (structural detail), but at higher dopant concentrations show some appreciable variation in scale. The tilt dependence of the $k$-factors of \cref{eq:k_factor_general} necessitates simulation. If the dopant concentration is relatively low, where the $k$-factor's dependence on dopant concentration is weak, $k$-factors simulated assuming any dopant concentration can be used for other doping concentrations, clearly a significant advantage when seeking to use simulation to determine an unknown dopant concentration. At higher dopant concentrations, an iterative process could be envisaged to refine the dopant concentration from an initial estimate. 

\begin{figure*}
    \centering
    \includegraphics[width=1\linewidth]{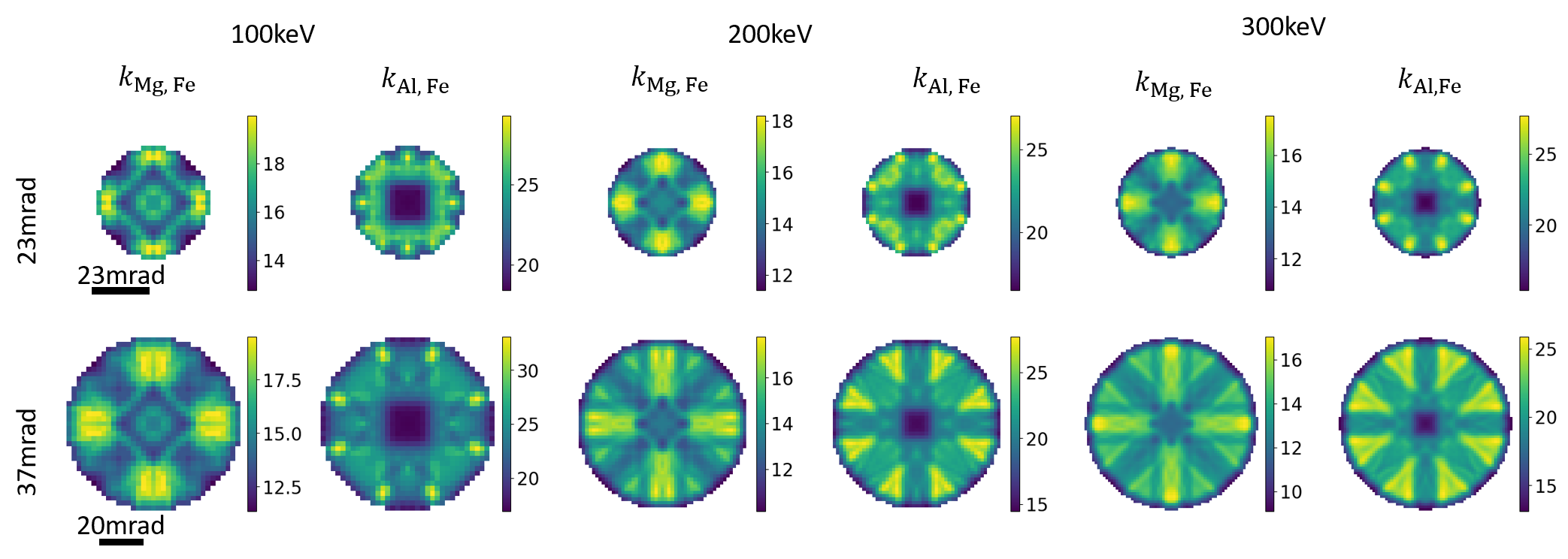}
    \caption{Tilt-dependent $k$-factors for an iron-doped spinel with 5\% on magnesium sites and 7\% on aluminium sites over a range of probe-forming aperture opening semi-angles and accelerating voltages, assuming the Fe-L$_{2,3}$-edge and the Mg and Al-K shells. All calculations assume $\varepsilon=10$eV.} 
    \label{fig:tilt_dep_k_factors}
\end{figure*}

To further explore what the tilt-dependent $k$-factors are sensitive to, \cref{fig:tilt_dep_k_factors} shows both $k_{\rm Mg, Fe}(\vec \theta)$ and $k_{\rm Al, Fe}(\vec \theta)$ for various accelerating voltages and probe-forming aperture angles. The site dependence is evident in the differences between $k_{\rm Mg, Fe}(\vec \theta)$ and $k_{\rm Al, Fe}(\vec \theta)$ when other parameters are kept constant. The tilt-dependence is clearly energy dependent, but this energy variation does not reduce the tilt-dependence of these $k$-factors, consistent with the tilt-independent analysis of \cref{fig:hn0_ALCHEMI_variation}(d)-(f) giving consistently poor concentration determinations at all these accelerating voltages. In \cref{fig:tilt_dep_k_factors}, the tilt-dependent $k$-factors are only shown for an angle range 10\% larger than the probe-forming aperture angle (since in the dark field region beyond the bright-field disk the signal is much lower, as seen in \cref{fig:ICPs_showcase}(c)). Thus the 37mrad case extends further, and so might be expected to better over-determine the ALCHEMI problem. (That the 23mrad case is similar to the inner 23mrad portion of the 37mrad case can perhaps more easily be understood via the traditional geometry of plane-wave incidence in \cref{fig:ICPs_showcase}(d): in that case, the tilt range and corresponding channelling conditions are then identical, and all that differs is the extent of the detector.)

\section{Experimental Considerations}
\subsection{Validation of Fractional Occupancy Approximation}\label{sec:frac_occ_justification}

\begin{figure*}[!t]%
\centering
\includegraphics[width=1\linewidth]{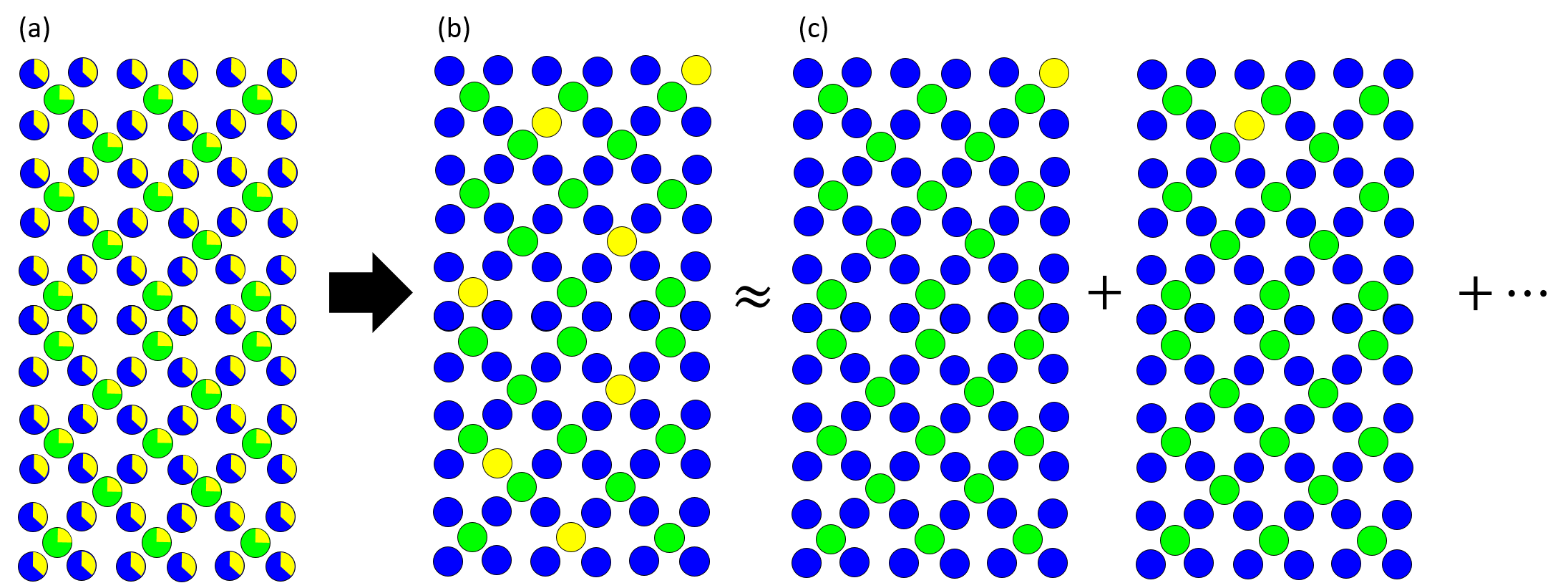}
\caption{Simulation methodology for checking the validity of the fractional occupancy model for performing EELS ALCHEMI on core-loss PACBED patterns. (a) Schematic of a fractional occupancy calculation, with some proportion of dopants (yellow) on all host sites (blue and green). (b) Schematic of a particular dopant configuration, where the dopant atoms (yellow) substitute for the host atoms (blue and green) at specific sites. (c) Two single-dopant configurations --- the ICPs from these two dopant configurations, along with others not shown, can be used to approximate the ICP for the specific dopant configuration in (b).}\label{fig:doping_method}
\end{figure*}

To this point, all core-loss scattering calculations of doped crystals presented have been based on the fractional occupancy model. As depicted in \cref{fig:doping_method}(a), in this model the elastic potential at every site along a dopant-containing column is set equal to the concentration-weighted average of the potentials of the dopant and host species (\cref{eq:fracocc_elastic}), and a similar weighting is applied to the inelastic intensities calculated using transition potentials (\cref{eq:total_intensity_fracocc}). However, the real situation is that depicted in \cref{fig:doping_method}(b): each site is either occupied by one species or the other. In this section we explore the validity of the fractional occupancy approximation for performing EELS ALCHEMI on core-loss PACBED patterns. 

As we will presently show, regions with the same average dopant concentration but different configurations of dopants along the columns can have somewhat different ICPs and so yield somewhat different dopant concentration estimates when analysed using (tilt-dependent) $k$-factors. However, more accurate simulations using fixed dopant configurations as per \cref{fig:doping_method}(b) are too time-consuming to allow a thorough statistical exploration of this variability. To explore a wide range of dopant configurations at relatively low dopant concentrations, we therefore introduce an alternative approximation. In this approximation we neglect the impact of doping on the scattering matrices, instead calculating them assuming an undoped crystal. For each site the dopant may occupy, we calculate the inelastic diffraction pattern from a dopant at that site by propagating the elastic wavefield through an undoped crystal up to that depth, accounting for the elastic and inelastic scattering from the dopant at that depth, and propagating the resultant inelastic wavefield through the remaining undoped crystal. We then approximate the ICP of any specific configuration of dopants by the sum of the single-dopant ICPs resulting from each of the constituent dopants. \Cref{fig:doping_method}(c) depicts two such single-dopant configurations.

We therefore now have three models. The first is the fractional occupancy model of \cref{fig:doping_method}(a). The second is a full calculation based on a particular configuration as per \cref{fig:doping_method}(b), in which a set of dopant locations is chosen (of number consistent with the desired dopant concentration) and both the elastic and inelastic scattering correctly includes that distribution of dopant atoms.\footnote{The $S$ operator approach discussed in connection with \cref{fig:calc_demo} is less suited to this case, and so in evaluating this model the multislice method was used instead.} We will call this the genuine doping model. The third is the approximation strategy of \cref{fig:doping_method}(c), which we will call the pseudo-doping model. The pseudo-doping model is approximate, and it is not \emph{a priori} clear how its accuracy compares with that of the other two models. However, we will explore this model next as it enables an efficient exploration of the variability that arises from different configurations.

\begin{figure*}[!t]%
    \centering
    \includegraphics[width=1\linewidth]{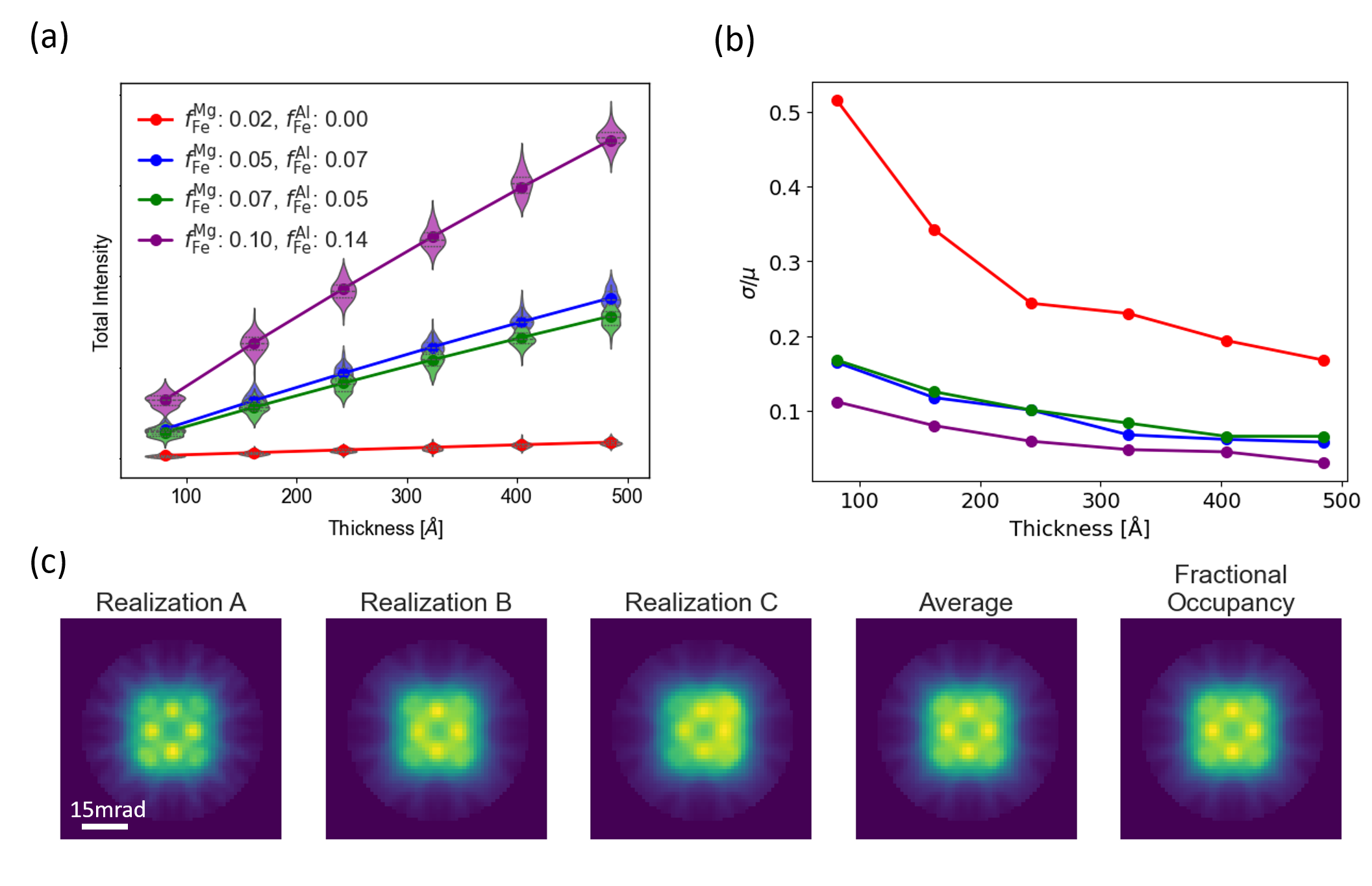}
    \caption{Comparison of core-loss diffraction as calculated using fractional occupancy against that calculated for various specific configurations of an iron-doped spinel assuming the Fe-L$_{2,3}$-edge and $\varepsilon=10$eV. (a) Plot of the integrated EELS intensity as a function of sample thickness for various dopant concentrations. The violin plots show the distribution of intensities of 1000 different doping configurations, and their mean is close to the fractional occupancy results (denoted by the dots joined by guide-to-the-eye lines).  (b) Plot of the standard deviation in integrated intensity scaled by the mean, showing the variations to be greater for lower doping concentrations and thicknesses. (c) Examples of iron L$_{2,3}$ shell core-loss PACBEDs assuming a 32nm thick sample with doping concentrations $f^{\rm Mg}_{\rm Fe}= 0.02, f^{\rm Al}_{\rm Fe}= 0.00$. The first three tiles show three random variations from a set of ten that were averaged to produce the fourth tile. The fifth (rightmost) tile corresponds to the PACBED calculated using fractional occupancy under the same conditions. All calculations use a 15.7mrad probe-forming aperture, a 30mrad detector and 300keV electrons.}
    \label{fig:frac_occ_comp}
\end{figure*}

A comparison of calculations using the fractional occupancy and pseudo-doping models is presented in \cref{fig:frac_occ_comp} for an iron-doped spinel. In all cases, the results represent the probe-position-averaged signal over a single unit cell. \Cref{fig:frac_occ_comp}(a) plots the integrated EELS intensity ({i.e.}~the total intensity in the ICP), for various dopant concentrations and sample thicknesses. The fractional occupancy calculation results are shown by the circles (joined by the solid line as a guide to the eye). The violin plots, calculated using the pseudo-doping model, visualise the variability in integrated EELS intensity across various configurations of dopants for the same overall concentrations. The fractional occupancy calculations are close to the average of the distributions for the various configurations in the pseudo-doping model, showing that fractional occupancy is a good approximation to the average signal across multiple configurations of dopants in the pseduo-doping model. However, the latter makes clear that the signal from specific configurations may appreciably fluctuate about that average. In particular, the considerable overlap of the violin plots from the $f_{\rm Fe}^{\rm Mg}=0.05$, $f_{\rm Fe}^{\rm Al}=0.07$ dopant concentration and the $f_{\rm Fe}^{\rm Mg}=0.07$, $f_{\rm Fe}^{\rm Al}=0.05$ dopant concentration means that, when averaging only over single units cells, we could not reliably distinguish those two different dopant distributions since their average difference is smaller than differences that can arise within either dopant concentration due to varying configurations of dopants along the columns. \Cref{fig:frac_occ_comp}(b) shows that the variability, measured by the ratio of the standard deviation of the distribution over configurations to its mean, decreases with increasing concentration and thickness, consistent with the increased likelihood of the average concentration being realised within the scan region under those conditions.

For a 32nm thick sample, the first three tiles of \cref{fig:frac_occ_comp}(c) show core-loss-filtered PACBED patterns (again averaging over only one unit cell) for three different doping configurations. While these patterns are broadly similar, some variability due to the difference in dopant configuration is evident. When patterns from ten different dopant configurations are averaged, the resultant diffraction pattern (fourth tile) is more symmetric and in good agreement with the result obtained from fractional occupancy (fifth tile). 

\begin{figure*}
     \centering
     \includegraphics[width=1\linewidth]{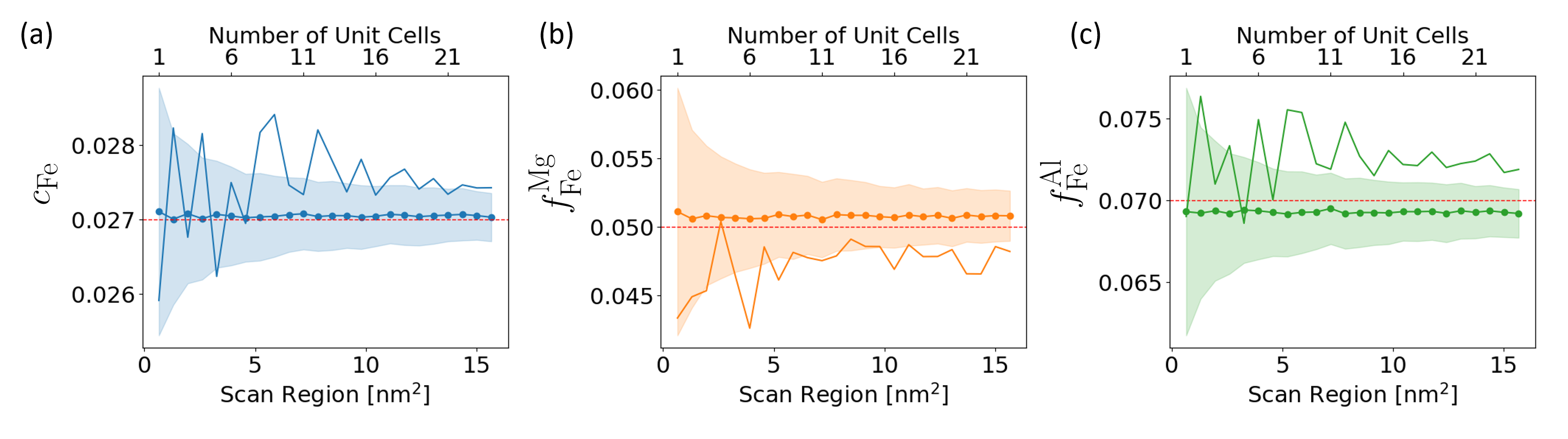}
     \caption{Dopant concentrations for (a) all sites, (b) Mg sites and (c) Al sites applying statistical EELS ALCHEMI using fractional occupancy $k$-factors to (infinite dose) core-loss PACBED ICPs for the Mg-K, Al-K and Fe-L$_{2,3}$ shell (from a spinel doped with 5\% Fe on Mg sites and 7\% Fe on Al sites) as a function of scan region. The dot-and-line shows the mean result from ICP calculations using the pseudo-doping model (\cref{fig:doping_method}(c)), with the shading denoting the region within one standard deviation of that mean (both calculated from 1000 doping configuration realizations). The true (input) concentrations are shown by the horizontal dashed red lines. The solid line results from ICP calculations using the genuine doping model (\cref{fig:doping_method}(b)) for a single doping configuration realization. As the scan region increases, the genuine doping result appears to be converging to a value that differs a little from the true concentration: fractional occupancy is an approximation, but a fairly good one. All calculations assume a 30mrad probe forming aperture, 300keV electrons and $\varepsilon=10$eV.}
     \label{fig:mixing_models_solves}
 \end{figure*}

As a function of the size of the scan region averaged over, \cref{fig:mixing_models_solves} compares concentrations determined from the genuine doping model and the pseudo-doping model, both analysed using (tilt-dependent) $k$-factors calculated via the fractional occupancy model. The genuine doping model is shown by the solid line, and because these calculations are very time-consuming only a single realisation of a dopant configuration is shown.\footnote{Again for computing time reasons, this calculation is actually an average over several smaller supercells, not an average over one single (prohibitively large) supercell.} For the pseudo-doping model, the dot-and-line plot shows the mean over many possible dopant configurations (specifically, 1000 distinct realisations) consistent with the overall average composition, and the shading shows the region within one standard deviation of that mean. \Cref{fig:mixing_models_solves}(a) shows the doping concentration for all sites, $c_{\rm Fe}$, \cref{fig:mixing_models_solves}(b) shows the fractional occupancy on Mg columns $f^{\rm Mg}_{\rm Fe}$, and \cref{fig:mixing_models_solves}(c) shows the fractional occupancy on Al columns $f^{\rm Al}_{\rm Fe}$, with the input (``ground truth'') values indicated by the red horizontal dashed line.

In the pseudo-doping model, the standard deviation decreases (the shaded regions narrow) with increasing scan region: consistent with \cref{fig:frac_occ_comp}(c), scan averaging over multiple configurations reduces the variability of the ICPs and so the concentrations determined. Averaging over a larger scan region of uniform concentration would thus be expected to increase the precision of the composition determination. (Non-uniform composition could only reliably be identified if the change in deduced concentration was larger than could be accounted for by variability with configuration of atoms along the column.) Consistent with \cref{fig:frac_occ_comp}(a), the mean of the concentrations deduced by analysing pseudo-doping model ICPs with the fractional occupancy model $k$-factors is in good agreement with the input composition. However, as both models are approximations, this is not the pertinent comparison for assessing the accuracy of the fractional occupancy model analysis. For that, we turn to the genuine doping model. 

Initially, the variability in the composition determined from genuine doping model ICPs is largely within the range of variability from the actual composition given by the pseudo-doping model. The variability in the genuine doping model is also seen to reduce with increasing scan area size. However, that model appears to be converging towards values that are offset somewhat from the true (input) concentration. This offset depends on the concentration assumed in the fractional occupancy model (remember, though relatively weak at lower concentrations, as per \cref{fig:k_fac_v_conc} the tilt-dependent $k$-factor varies somewhat with concentration assumed). We do not suggest that assuming the ``right'' concentration when calculating the $k$-factor would give especially favourable results. Our purpose is rather to acknowledge that fractional occupancy is an approximation, and some discrepancy is to be expected. That said, the discrepancies between the genuine doping analysis and the actual concentration in \cref{fig:mixing_models_solves} are relatively small. As we next show, use of the fractional occupancy approximation is unlikely to be the dominant limitation in practice.

\subsection{Noise Analysis}\label{sec:noise_analysis}

\begin{figure*}
    \centering
    \includegraphics[width=1\linewidth]{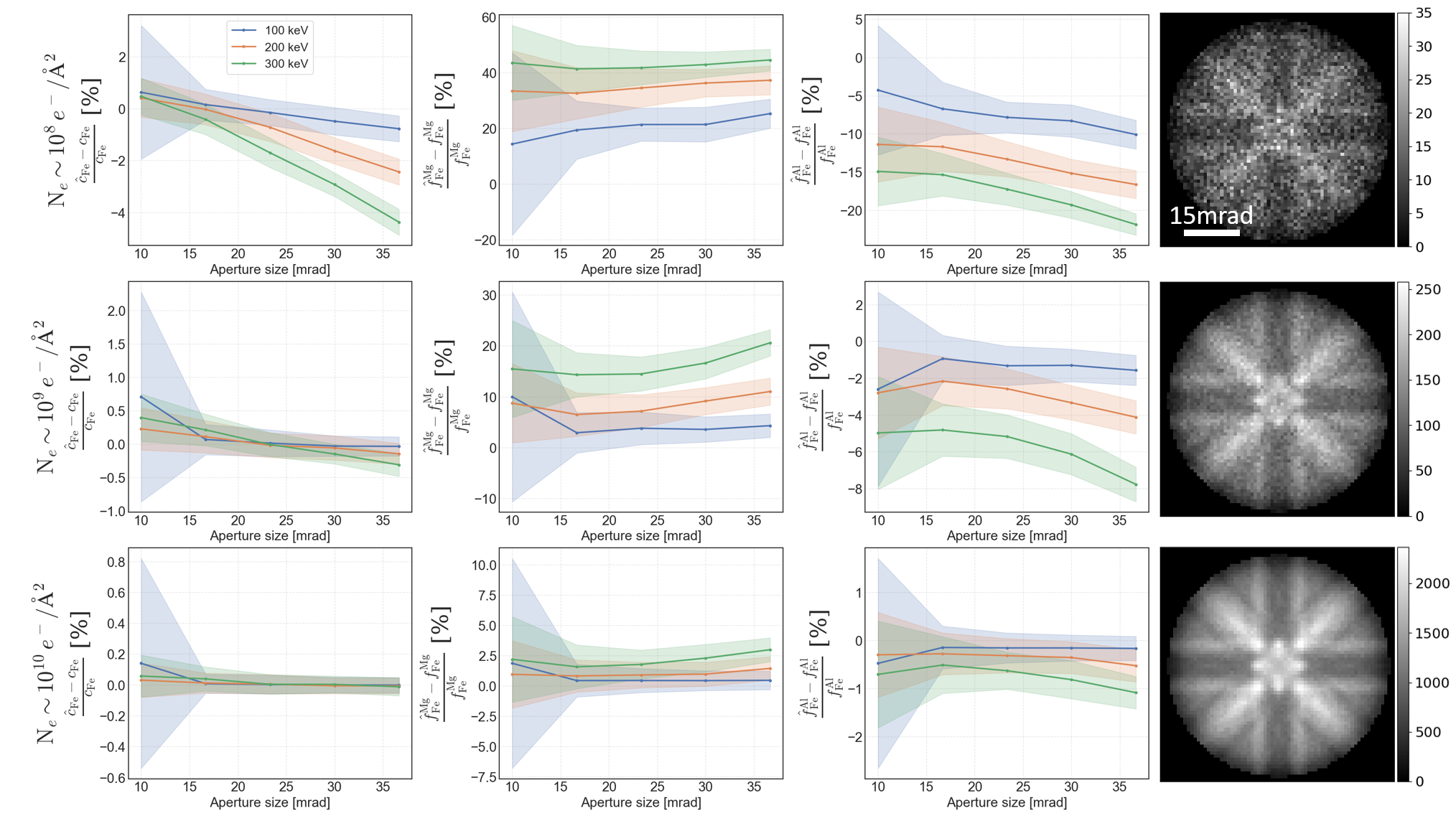}
    \caption{Average percentage error in dopant concentration in iron-doped spinel from simulated ICPs that include shot noise for the dose levels shown at the left of each row. The dose increases an order of magnitude across successive rows. Along the columns, from left to right, is the overall dopant concentration, the concentration on Mg and Al sites, and an example Mg-K ICP at that dose. All ICPs were simulated at 10eV beyond the corresponding shell edge, and a 10eV energy window assumed by approximating the integral by a single rectangle of width 10eV. All ICPs simulated were averaged over just a single unit cell.}
    \label{fig:noise_error}
\end{figure*}

Thus far, all simulations have assumed noise-free ICPs, essentially assuming infinite dose. In practice, the low cross-section for ionization combined with distributing that signal across multiple pixels on the 4D STEM detector means that experimental patterns will likely be noisy. There are a few ways to reduce the noise in the data. First, we should be judicious when choosing which edges to use: the edges chosen for the ICPs have a profound effect on the dose required to obtain ICPs with sufficient signal-to-noise ratio for EELS ALCHEMI analysis. For the present spinel sample, the Mg- and Al-L shell edges are not experimentally detectable (above other low loss scattering mechanisms), obliging us to work with the K shell edges, despite the Mg-K and Al-K edges (1305eV and 1560eV, respectively) having a low cross-section. However, for the Fe dopants, we consider the L$_{2,3}$-edge, which is an order of magnitude more intense than the K shell edges for Mg and Al. Second, if spatial resolution is not a priority, then we should scan over the widest region that we expect to have roughly uniform thickness and concentration, increasing the total signal without increasing the dose. As per the previous section, this also helps by meaning the $k$-factors calculated assuming fractional occupancy better match the experimental ICPs. Third, the energy window should be as wide as possible, provided the background subtraction is reliable and the delocalisation of the transition potentials remains constant across the entire window. This requires the range of energies to be insignificant in comparison with the edge energy, which is usually a non-issue except for very light elements, say below oxygen. 

The average percentage error for an iron-doped spinel is shown in \cref{fig:noise_error} for increasing levels of dose in each rows and using tilt-dependent $k$-factors. The percentage error is plotted as a function of aperture opening semi-angle for several accelerating voltages. We stress that the dose provided assumes scanning over just a single unit cell --- if multiple unit cells are averaged over then the necessary dose can be correspondingly reduced. To visualize the impact of dose on the ICPs we also show the Mg-K ICP, since it contains the greatest structure. 

As noted earlier, $c_{\rm Fe}$ has the smallest errors. As such, if only the overall dopant concentration is required, a dose on the order of $10^8-10^9~e^-/{\rm \AA}^2$ is required (if only a single unit cell is scanned over, but the necessary dose could be lowered by scanning over multiple unit cells). Alternatively, to obtain accurate estimates of the partitioning of dopants across crystallographic sites, a dose of at least $10^{10}~e^-/{\rm \AA}^2$ is required (which again could be lowered by scanning over multiple unit cells). Observe that the percentage error drops an order of magnitude as the dose is increased by an order of magnitude, providing a rough rule of thumb.

\section{Conclusion}
We have utilized the reciprocity between STEM and TEM, which holds to a good approximation in core-loss electron scattering, to relate core-loss PACBED patterns to rocking patterns resulting from a plane-wave 2D tilt series recorded with a bucket detector. Given the close resemblance of this approach to ALCHEMI, we have argued that core-loss PACBED patterns are approximately linearly dependent. This linear dependence can be leveraged to infer dopant concentrations and distributions, analogous to EDX ALCHEMI.  

The accuracy of EELS ALCHEMI was found to be sensitive to differences in the inelastic interaction ranges of the constituent atoms. This issue, while also present in EDX ALCHEMI, is significantly more pronounced in EELS ALCHEMI. In some cases, this can be partly mitigated by increasing the accelerating voltage and using very small or large opening angles, say $\sim10$mrad and $\sim60$mrad respectively. However, to achieve reasonable results in any experimental configuration, we propose a generalization of the $k$-factors that can only be obtained though detailed simulations. In theory, this approach provides exact results, but in practice it requires careful consideration of the validity of the fractional occupancy approximation and an adequate electron dose. For the spinel studied here, the required dose would be on the order of $10^8-10^{10}~e^-/{\rm \AA}^2$ if single unit cell resolution was sought, but would be reduced accordingly if spatial resolution were traded for improved signal-to-noise ratio by averaging over multiple unit cells.

\section{Competing interests}
No competing interest is declared.

\section{Author contributions statement}
M.D.: Writing – review \& editing, Writing – original draft, Visualization, Software, Methodology, Investigation, Formal
analysis. T.C.P.: Writing – review \& editing, Supervision, Methodology, Conceptualization. M.W.: Writing – review \& editing, Supervision, Methodology, Conceptualization. S.D.F.: Writing – review \& editing, Validation, Methodology, Funding acquisition, Conceptualization.

\section{Acknowledgments}
The authors thank Dr Peter Miller for helpful discussions. This research is supported by an Australian Government Research Training Program (RTP) Scholarship. SDF is the recipient of an Australian Research Council Future Fellowship (project number FT190100619) funded by the Australian Government.

\bibliographystyle{abbrvnat}
\bibliography{references_without_doi}

\begin{thebibliography}{57}
\providecommand{\natexlab}[1]{#1}
\providecommand{\url}[1]{\texttt{#1}}
\expandafter\ifx\csname urlstyle\endcsname\relax
  \providecommand{\doi}[1]{doi: #1}\else
  \providecommand{\doi}{doi: \begingroup \urlstyle{rm}\Url}\fi

\bibitem[Anderson(1997)]{anderson1997alchemi}
I.~M. Anderson.
\newblock {ALCHEMI} study of site distributions of {3D}-transition metals in {B}2-ordered iron aluminides.
\newblock \emph{Acta Mater}, 45:\penalty0 3897--3909, 1997.

\bibitem[Blom and Vogt(2020)]{blom2020probing}
D.~A. Blom and T.~Vogt.
\newblock Probing compositional order in atomic columns: {STEM} simulations beyond the virtual crystal approximation.
\newblock \emph{Microsc Microanal}, 26:\penalty0 46--52, 2020.

\bibitem[Brown et~al.(2020)Brown, Pelz, Ophus, and Ciston]{10.1017/S1431927620023326}
H.~Brown, P.~Pelz, C.~Ophus, and J.~Ciston.
\newblock A python based open-source multislice simulation package for transmission electron microscopy.
\newblock \emph{Microsc Microanal}, 26S2:\penalty0 2954--2956, 2020.

\bibitem[Brown(2025)]{brown2025py_multislice_repo}
H.~G. Brown.
\newblock py\_multislice, 2025.
\newblock URL \url{https://github.com/HamishGBrown/py_multislice.git}.

\bibitem[Brown et~al.(2019)Brown, Ciston, and Ophus]{PhysRevResearch.1.033186}
H.~G. Brown, J.~Ciston, and C.~Ophus.
\newblock Linear-scaling algorithm for rapid computation of inelastic transitions in the presence of multiple electron scattering.
\newblock \emph{Phys Rev Res}, 1:\penalty0 033186, 2019.

\bibitem[Buseck and Self(1992)]{buseck1992electron}
P.~R. Buseck and P.~Self.
\newblock Electron energy-loss spectroscopy ({EELS}) and electron channelling ({ALCHEMI}).
\newblock \emph{Reviews in Mineralogy}, 27:\penalty0 141--180, 1992.

\bibitem[Bustillo et~al.(2021)Bustillo, Zeltmann, Chen, Donohue, Ciston, Ophus, and Minor]{bustillo20214d}
K.~C. Bustillo, S.~E. Zeltmann, M.~Chen, J.~Donohue, J.~Ciston, C.~Ophus, and A.~M. Minor.
\newblock {4D}-{STEM} of beam-sensitive materials.
\newblock \emph{Acc Chem Res}, 54:\penalty0 2543--2551, 2021.

\bibitem[Carlino and Grillo(2005)]{carlino2005atomic}
E.~Carlino and V.~Grillo.
\newblock Atomic-resolution quantitative composition analysis using scanning transmission electron microscopy {Z}-contrast experiments.
\newblock \emph{Phys Rev B}, 71:\penalty0 235303, 2005.

\bibitem[Chejarla et~al.(2023)Chejarla, Ahmed, Belz, Scheunert, Beyer, and Volz]{chejarla2023measuring}
V.~S. Chejarla, S.~Ahmed, J.~Belz, J.~Scheunert, A.~Beyer, and K.~Volz.
\newblock Measuring spatially-resolved potential drops at semiconductor hetero-interfaces using {4D}-{STEM}.
\newblock \emph{Small Methods}, 7:\penalty0 2300453, 2023.

\bibitem[Cliff and Lorimer(1975)]{cliff1975quantitative}
G.~Cliff and G.~W. Lorimer.
\newblock The quantitative analysis of thin specimens.
\newblock \emph{J Microsc}, 103:\penalty0 203--207, 1975.

\bibitem[Coene and {Van Dyck}(1990)]{coene_inelastic_1990}
W.~Coene and D.~{Van Dyck}.
\newblock Inelastic scattering of high-energy electrons in real space.
\newblock \emph{Ultramicroscopy}, 33:\penalty0 261--267, 1990.

\bibitem[{Cowley}(1969)]{cowley_1969}
J.~M. {Cowley}.
\newblock {Image Contrast in a Transmission Scanning Electron Microscope}.
\newblock \emph{Appl Phys Lett}, 15:\penalty0 58--59, 1969.

\bibitem[Deimetry et~al.(2024)Deimetry, Petersen, Brown, Weyland, and Findlay]{DEIMETRY2024114036}
M.~Deimetry, T.~C. Petersen, H.~G. Brown, M.~Weyland, and S.~D. Findlay.
\newblock Differential phase contrast from electrons that cause inner shell ionization.
\newblock \emph{Ultramicroscopy}, 266:\penalty0 114036, 2024.

\bibitem[Dwyer(2005)]{dwyer_multislice_2005}
C.~Dwyer.
\newblock Multislice theory of fast electron scattering incorporating atomic inner-shell ionization.
\newblock \emph{Ultramicroscopy}, 104:\penalty0 141--151, 2005.

\bibitem[Dwyer et~al.(2008)Dwyer, Findlay, and Allen]{PhysRevB.77.184107}
C.~Dwyer, S.~D. Findlay, and L.~J. Allen.
\newblock Multiple elastic scattering of core-loss electrons in atomic resolution imaging.
\newblock \emph{Phys Rev B}, 77:\penalty0 184107, 2008.

\bibitem[D’Alfonso et~al.(2008)D’Alfonso, Cosgriff, Findlay, Behan, Kirkland, Nellist, and Allen]{dalfonso_three-dimensional_2008}
A.~D’Alfonso, E.~Cosgriff, S.~Findlay, G.~Behan, A.~Kirkland, P.~Nellist, and L.~Allen.
\newblock Three-dimensional imaging in double aberration-corrected scanning confocal electron microscopy, part {II}: {I}nelastic scattering.
\newblock \emph{Ultramicroscopy}, 108:\penalty0 1567--1578, 2008.

\bibitem[Findlay et~al.(2007)Findlay, Schattschneider, and Allen]{FINDLAY200758}
S.~D. Findlay, P.~Schattschneider, and L.~J. Allen.
\newblock Imaging using inelastically scattered electrons in {CTEM} and {STEM} geometry.
\newblock \emph{Ultramicroscopy}, 108:\penalty0 58--67, 2007.

\bibitem[Findlay et~al.(2008)Findlay, Oxley, and Allen]{findlay2008modeling}
S.~D. Findlay, M.~P. Oxley, and L.~J. Allen.
\newblock Modeling atomic-resolution scanning transmission electron microscopy images.
\newblock \emph{Microsc Microanal}, 14:\penalty0 48--59, 2008.

\bibitem[Gebhart et~al.(2025)Gebhart, Schretter, Krapf, Merle, Cordill, and Gammer]{gebhart2025grain}
D.~D. Gebhart, L.~Schretter, A.~Krapf, B.~Merle, M.~J. Cordill, and C.~Gammer.
\newblock Grain rotation and crack propagation in bulge-tested gold films with {4D}-{STEM}.
\newblock \emph{JOM}, pages 1--10, 2025.

\bibitem[Haas and Koch(2022)]{hass_koch}
B.~Haas and C.~T. Koch.
\newblock Momentum- and energy-resolved {STEM} at atomic resolution.
\newblock \emph{Microsc Microanal}, 28S1:\penalty0 406--408, 2022.

\bibitem[Jones(2003)]{jones2003determining}
I.~P. Jones.
\newblock Determining the locations of chemical species in ordered compounds: {ALCHEMI}.
\newblock In \emph{Adv Imag Electron Phys}, volume 125, pages 63--I. 2003.

\bibitem[Joy et~al.(1982)Joy, Newbury, and Davidson]{joy1982electron}
D.~C. Joy, D.~E. Newbury, and D.~L. Davidson.
\newblock Electron channeling patterns in the scanning electron microscope.
\newblock \emph{J Appl Phys}, 53\penalty0 (8):\penalty0 R81--R122, 1982.

\bibitem[Kohl and Rose(1985)]{KOHL1985173}
H.~Kohl and H.~Rose.
\newblock Theory of image formation by inelastically scattered electrons in the electron microscope.
\newblock volume~65 of \emph{Adv Electron Electron Phys}, pages 173--227. Academic Press, 1985.

\bibitem[Krause and Rosenauer(2017)]{KRAUSE20171}
F.~F. Krause and A.~Rosenauer.
\newblock Reciprocity relations in transmission electron microscopy: {A} rigorous derivation.
\newblock \emph{Micron}, 92:\penalty0 1--5, 2017.

\bibitem[LeBeau et~al.(2010)LeBeau, Findlay, Allen, and Stemmer]{PACBED_lebeau}
J.~M. LeBeau, S.~D. Findlay, L.~J. Allen, and S.~Stemmer.
\newblock Position averaged convergent beam electron diffraction: {T}heory and applications.
\newblock \emph{Ultramicroscopy}, 110:\penalty0 118--125, 2010.

\bibitem[Lee et~al.(2001)Lee, Kumada, and Yoshio]{LEE2001376}
Y.-S. Lee, N.~Kumada, and M.~Yoshio.
\newblock Synthesis and characterization of lithium aluminum-doped spinel ({LiAl$_x$Mn$_{2-x}$O$_4$}) for lithium secondary battery.
\newblock \emph{J Power Sources}, 96:\penalty0 376--384, 2001.

\bibitem[Li et~al.(2024)Li, Biskupek, Linck, Rose, K{\"u}kelhan, M{\"u}ller, and Kaiser]{li2024efficient}
Z.~Li, J.~Biskupek, M.~Linck, H.~Rose, P.~K{\"u}kelhan, H.~M{\"u}ller, and U.~Kaiser.
\newblock An efficient electron ptychography method for retrieving the object spectrum from only a few iterations.
\newblock \emph{Microsc Microanal}, 30:\penalty0 294--305, 2024.

\bibitem[L{\"o}ffler and Ederer(2023)]{loffler20234d}
S.~L{\"o}ffler and M.~Ederer.
\newblock {4D} energy-filtered {STEM}: A new approach for mapping orbital transitions.
\newblock \emph{Microsc Microanal}, 29S1:\penalty0 376, 2023.

\bibitem[Madsen and Susi(2021)]{madsen_abtem_2021}
J.~Madsen and T.~Susi.
\newblock {{abTEM}}: Transmission electron microscopy from first principles.
\newblock \emph{Open Research Europe}, 1\penalty0 (24):\penalty0 13015, 2021.

\bibitem[Madsen and Susi(2025)]{abtem2025abTEM}
J.~Madsen and T.~Susi.
\newblock {abTEM}, 2025.
\newblock URL \url{https://github.com/abTEM/abTEM.git}.

\bibitem[Mariosi et~al.(2020)Mariosi, Venturini, {da Cas Viegas}, and Bergmann]{MARIOSI20202772}
F.~R. Mariosi, J.~Venturini, A.~{da Cas Viegas}, and C.~P. Bergmann.
\newblock Lanthanum-doped spinel cobalt ferrite ({CoFe$_2$O$_4$}) nanoparticles for environmental applications.
\newblock \emph{Ceram Int}, 46:\penalty0 2772--2779, 2020.

\bibitem[Matsumura et~al.(1999)Matsumura, Soeda, Zaluzec, and Kinoshita]{matsumura1999electron}
S.~Matsumura, T.~Soeda, N.~J. Zaluzec, and C.~Kinoshita.
\newblock Electron channeling {X}-ray microanalysis for cation configuration in irradiated magnesium aluminate spinel.
\newblock \emph{MRS Proceedings}, 589:\penalty0 129--134, 1999.

\bibitem[Mellini and Menichini(1985)]{mellini1985proportionality}
H.~Mellini and R.~Menichini.
\newblock Proportionality factors for thin film {TEM/EDS} microanalysis of silicate minerals.
\newblock \emph{Rend Soc Ital Mineral Petrol}, 40:\penalty0 261--266, 1985.

\bibitem[Midgley et~al.(1995)Midgley, Saunders, Vincent, and Steeds]{midgley1995energy}
P.~A. Midgley, M.~Saunders, R.~Vincent, and J.~W. Steeds.
\newblock Energy-filtered convergent-beam diffraction: examples and future prospects.
\newblock \emph{Ultramicroscopy}, 59:\penalty0 1--13, 1995.

\bibitem[Morgan(1973)]{morgan1973}
D.~B. Morgan.
\newblock \emph{Channeling : theory, observation and applications}.
\newblock Wiley, London, New York, 1973.

\bibitem[M{\"u}ller-Caspary et~al.(2017)M{\"u}ller-Caspary, Krause, Grieb, L{\"o}ffler, Schowalter, B{\'e}ch{\'e}, Galioit, Marquardt, Zweck, Schattschneider, et~al.]{muller2017measurement}
K.~M{\"u}ller-Caspary, F.~F. Krause, T.~Grieb, S.~L{\"o}ffler, M.~Schowalter, A.~B{\'e}ch{\'e}, V.~Galioit, D.~Marquardt, J.~Zweck, P.~Schattschneider, et~al.
\newblock Measurement of atomic electric fields and charge densities from average momentum transfers using scanning transmission electron microscopy.
\newblock \emph{Ultramicroscopy}, 178:\penalty0 62--80, 2017.

\bibitem[Muto and Ohtsuka(2017)]{muto2017high}
S.~Muto and M.~Ohtsuka.
\newblock High-precision quantitative atomic-site-analysis of functional dopants in crystalline materials by electron-channelling-enhanced microanalysis.
\newblock \emph{Prog Cryst Growth Charact Mater}, 63:\penalty0 40--61, 2017.

\bibitem[Ofori et~al.(2005)Ofori, Rossouw, and Humphreys]{ofori2005determining}
A.~P. Ofori, C.~J. Rossouw, and C.~J. Humphreys.
\newblock Determining the site occupancy of {Ru} in the {L1$_2$} phase of a {Ni}-base superalloy using {ALCHEMI}.
\newblock \emph{Acta Mater}, 53:\penalty0 97--110, 2005.

\bibitem[Ohishi et~al.(2002)Ohishi, Ogawa, Watanabe, Katsumata, Komuro, Morikawa, and Toba]{aizawa2002characteristics}
N.~Ohishi, S.~Ogawa, E.~Watanabe, T.~Katsumata, S.~Komuro, T.~Morikawa, and E.~Toba.
\newblock Characteristics of chromium doped spinel crystals for a fiber-optic thermometer application.
\newblock \emph{Rev Sci Instrum}, 73:\penalty0 3089--3092, 2002.

\bibitem[Ohtsuka et~al.(2021)Ohtsuka, Oda, Tanaka, Kitaoka, and Muto]{ohtsuka20212d}
M.~Ohtsuka, K.~Oda, M.~Tanaka, S.~Kitaoka, and S.~Muto.
\newblock {2D}-{HARECXS} analysis of dopant and oxygen vacancy sites in {Al}-doped yttrium titanate.
\newblock \emph{J Am Ceram Soc}, 104:\penalty0 3760--3769, 2021.

\bibitem[Ophus(2019)]{10.1017/S1431927619000497}
C.~Ophus.
\newblock Four-dimensional scanning transmission electron microscopy ({4D}-{STEM}): From scanning nanodiffraction to ptychography and beyond.
\newblock \emph{Microsc Microanal}, 25:\penalty0 563--582, 2019.

\bibitem[Oxley and Allen(2001)]{oxley2001atomic}
M.~P. Oxley and L.~J. Allen.
\newblock Atomic scattering factors for {K}-shell electron energy-loss spectroscopy.
\newblock \emph{Acta Crystallogr A}, 57:\penalty0 713--728, 2001.

\bibitem[Oxley et~al.(1999)Oxley, Allen, and Rossouw]{OXLEY1999109}
M.~P. Oxley, L.~J. Allen, and C.~J. Rossouw.
\newblock Correction terms and approximations for atom location by channelling enhanced microanalysis.
\newblock \emph{Ultramicroscopy}, 80:\penalty0 109--124, 1999.

\bibitem[Pennycook(1988)]{PENNYCOOK1988239}
S.~Pennycook.
\newblock Delocalization corrections for electron channeling analysis.
\newblock \emph{Ultramicroscopy}, 26:\penalty0 239--248, 1988.

\bibitem[Pogany and Turner(1968)]{pogany_turner}
A.~P. Pogany and P.~S. Turner.
\newblock Reciprocity in electron diffraction and microscopy.
\newblock \emph{Acta Crystallogr A}, 24:\penalty0 103--109, 1968.

\bibitem[Qi et~al.(2021)Qi, Li, Du, Shi, Huang, Yang, Liu, Xu, Dai, Yu, et~al.]{qi2021four}
R.~Qi, N.~Li, J.~Du, R.~Shi, Y.~Huang, X.~Yang, L.~Liu, Z.~Xu, Q.~Dai, D.~Yu, et~al.
\newblock Four-dimensional vibrational spectroscopy for nanoscale mapping of phonon dispersion in {BN} nanotubes.
\newblock \emph{Nat Commun}, 12:\penalty0 1179, 2021.

\bibitem[Robert et~al.(2022)Robert, Diederichs, and M{\"u}ller-Caspary]{robert2022contribution}
H.~Robert, B.~Diederichs, and K.~M{\"u}ller-Caspary.
\newblock Contribution of multiple plasmon scattering in low-angle electron diffraction investigated by energy-filtered atomically resolved {4D}-{STEM}.
\newblock \emph{Appl Phys Lett}, 121, 2022.

\bibitem[Rossouw(1995)]{rossouw1995incoherent}
C.~J. Rossouw.
\newblock Incoherent contrast under dynamical diffraction conditions.
\newblock \emph{Ultramicroscopy}, 58:\penalty0 211--222, 1995.

\bibitem[Rossouw et~al.(1989)Rossouw, Turner, White, and {O'Connor}]{rossouw_statistical_alchemi}
C.~J. Rossouw, P.~S. Turner, T.~J. White, and A.~J. {O'Connor}.
\newblock Statistical analysis of electron channelling microanalytical data for the determination of site occupancies of impurities.
\newblock \emph{Phil Mag Lett}, 60:\penalty0 225--232, 1989.

\bibitem[Rossouw et~al.(1996)Rossouw, Forwood, Gibson, and Miller]{rossouw1996statistical}
C.~J. Rossouw, C.~T. Forwood, M.~A. Gibson, and P.~R. Miller.
\newblock Statistical {ALCHEMI}: {A} general formulation and method with application to {Ti--Al} ternary alloys.
\newblock \emph{Philos Mag A}, 74:\penalty0 57--76, 1996.

\bibitem[Rossouw et~al.(1997)Rossouw, Forwood, Gibson, and Miller]{rossouw1997generation}
C.~J. Rossouw, C.~T. Forwood, M.~A. Gibson, and P.~R. Miller.
\newblock Generation and absorption of characteristic {X}-rays under dynamical electron diffraction conditions.
\newblock \emph{Micron}, 28:\penalty0 125--137, 1997.

\bibitem[Spence(1992)]{spence1992channelling}
J.~C.~H. Spence.
\newblock Electron channelling and its uses.
\newblock volume~1 of \emph{Electron Diffraction Techniques}, pages 465--532. Oxford University Press, 1992.

\bibitem[Spence and Taftø(1983)]{spence_alchemi_1983}
J.~C.~H. Spence and J.~Taftø.
\newblock {ALCHEMI}: a new technique for locating atoms in small crystals.
\newblock \emph{J Microsc}, 130:\penalty0 147--154, 1983.

\bibitem[Spence et~al.(1988)Spence, Kuwabara, and Kim]{spence1988localization}
J.~C.~H. Spence, M.~Kuwabara, and Y.~Kim.
\newblock Localization effects on quantification in axial and planar {ALCHEMI}.
\newblock \emph{Ultramicroscopy}, 26:\penalty0 103--112, 1988.

\bibitem[Walls(1992)]{walls1992formulation}
M.~G. Walls.
\newblock A formulation of {ALCHEMI} for materials containing light elements.
\newblock \emph{Microsc Microanal Microstruct}, 3:\penalty0 443--451, 1992.

\bibitem[Williams and Carter(1996)]{willans_carter}
D.~B. Williams and C.~B. Carter.
\newblock \emph{Transmission electron microscopy : a textbook for materials science}, chapter~35.
\newblock Plenum Press, New York, 1996.

\bibitem[Zaluzec et~al.(2005)Zaluzec, Blackford, Smith, and Colella]{zaluzec2005hareces}
N.~J. Zaluzec, M.~G. Blackford, K.~L. Smith, and M.~Colella.
\newblock {HARECES} measurements of carbon {K} shell excitation in graphite.
\newblock \emph{Microsc Microanal}, 11S2:\penalty0 718--719, 2005.

\end{thebibliography}

\end{document}